\documentclass[extra,mreferee]{gji}

\usepackage{amssymb,amsmath}
\usepackage{graphicx}
\usepackage[caption=false]{subfig}
\usepackage{bm}
\usepackage{physics}
\usepackage{wasysym}
\usepackage{color,soul}

\newcommand{\classname}[1]{\texttt{#1}}

\begin{document}

\title{The coupling between inertial and rotational eigenmodes in planets with liquid cores}

\author[S.~A.~Triana, J.~Rekier, A.~Trinh and V.~Dehant]{Santiago Andr\'es Triana$^1$, J\'er\'emy Rekier$^1$, Antony Trinh$^1$ and Veronique Dehant$^1$\\
$^1$ Royal Observatory of Belgium, Ringlaan 3, BE-1180 Brussels, Belgium}

\maketitle

\begin{summary}
The Earth is a rapidly rotating body. The centrifugal pull makes its shape resemble a flattened ellipsoid and Coriolis forces support waves in its fluid core, known as inertial waves. These waves can lead to global 
oscillations, or modes, of the fluid. Periodic variations of the Earth's rotation axis (nutations) can lead to an exchange of angular momentum between the mantle and the fluid core
 and excite these inertial modes. In addition to viscous torques that exist regardless of the shape of the boundaries, the small flattening of the core-mantle boundary (CMB) allows inertial modes to exert pressure torques on the mantle. These torques effectively couple the rigid-body dynamics of the Earth with the fluid dynamics of the fluid core. Here we present the first high resolution numerical model that solves simultaneously the rigid body dynamics of the mantle and the Navier-Stokes equation for the liquid core. This method takes naturally into account dissipative processes in the fluid that are ignored in current nutation models. We find that the Free Core Nutation (FCN) mode, mostly a toroidal fluid flow if the mantle has a large moment of inertia, enters into resonance with nearby modes if the mantle's moment of inertia is reduced. These mode interactions seem to be completely analogous to the ones discovered by \citet{schmitt2006} in a uniformly rotating ellipsoid  with varying flattening.    
\end{summary}


\section{\label{sec:intro} Introduction} 
The Very Long Baseline Interferometry (VLBI) technique allows us to establish with great accuracy the instantaneous spin axis of the Earth relative to an inertial frame built by observing radio signals from distant quasars. We also know with good accuracy the external torques acting on our spinning planet, mainly from the Sun and the Moon. Thus, knowing both the forcing and the planet's response, and assisted with other valuable data from seismology and geomagnetism, it is possible to infer reasonably accurate models of the Earth and its interior. This indeed is the case, see e.g. \citet{mathews2002}. The model's parameters, known as the Basic Earth Parameters (BEP), are obtained by fitting the VLBI observations to the models. Mantle elasticity, ocean loading and atmospheric effects make those models highly sophisticated and also successful in explaining most, but not all of the observations, for example as noted by  \citet{koot2010}. This might be related to the fact that the treatment of the liquid core flow is not nearly as sophisticated as other elements in the model. In fact, the core flow is modelled simply as an inviscid, incompressible flow with uniform vorticity, which for some specific purposes might be sufficient. On the other hand, there is a vast amount of experimental and numerical work on rapidly rotating fluids, where the motion of the bounding surface is prescribed (as steady rotation, including precession or libration), that has revealed surprising features and instabilities related to inertial modes (i.e. modes restored by the Coriolis force), see  \citet{lebars2015} for a review.

One of the problems of modelling planets with inviscid fluids is that although viscous effects might not be important a priori, the limit when viscosity tends to zero is not well behaved and inviscid solutions might become singular, a fact already noted by  \citet{rieutord2001}. In a spherical shell, like the Earth's fluid core, the only inertial modes that stay regular in the inviscid limit are the purely toroidal ones, also called planetary or Rossby modes. In astrophysical or geophysical applications the small but finite viscosity regularizes any singularities, which appear instead as detached shear layers  \citep{hollerbach1995,rieutord1997,rieutord2001}. There is consequently additional energy dissipation caused by these shear layers, but more importantly, in the case electrically conducting fluids, the ohmic dissipation is large in these layers and can become the main energy dissipation mechanism as the inviscid limit is approached, see  \citet{buffett2010} and  \citet{lin2017}.

Another aspect is the possible interaction between inertial modes and the rotational modes of the Earth as studied by  \citet{rogister2009}. This study is quite suggestive because it shows that avoided crossings can take place under certain conditions. However the limited resolution they could achieve precluded them to identify unambiguously their eigenmodes as inertial modes, apart from the fact that the fluid core is assumed inviscid, therefore limited to \emph{regular} (i.e. purely toroidal) inertial modes in their spherical shell configuration. The present study has a similar aim, namely to understand how the rotational modes like the Free Core Nutation (FCN) and the Chandler Wobble (CW) interact with inertial modes. 

As a first step towards a better fluid dynamical treatment of the problem we consider a rigid spheroidal mantle and a completely fluid core, including viscosity. Instead of considering eigendisplacements as done by  \citet{rogister2009} we consider the Navier-Stokes equation (linear in velocity) to model with high resolution the core flow, including the motion of the mantle through the Coriolis and Poincar\'e forces. Simultaneously, we compute the motion of the mantle by considering the torques exerted by the fluid on the mantle (Euler equations). To deal with the spheroidal shape of the CMB, we adopt the same Taylor expansion method that we developed in  \citet{rekier2018}. We discuss in detail the model in Section \ref{sec:model} and the numerical method in Section \ref{sec:num}. Sections \ref{sec:res} through \ref{sec:disc} are devoted to the presentation and discussion of the results. In Section \ref{sec:end} we end with a summary and perspectives for future work. The appendix presents important technical details.

\section{The coupled inertial-rotational model}
\label{sec:model}
We consider a two-layer model planet with a completely fluid core enclosed by a rigid, oblate and axisymmetric ellipsoidal mantle. The basic (unperturbed) state of the planet is that of uniform angular speed $\Omega_0$ throughout and around the short axis of symmetry (the $\hat {\bm z}$ axis). We allow the mantle's spin $\vb{\Omega}(t)$, as referred to an inertial frame, to respond freely to torques exerted by the fluid core. We assume small perturbed motions compared to the basic state. 

\subsection{The fluid core}
We model the core as an incompressible, homogeneous fluid with kinematic viscosity $\nu$ and density $\rho_f$. We describe the flow, $\vb{u}$, as observed from a reference frame attached rigidly to the mantle, where the boundary conditions can be prescribed more conveniently. With this choice, the dimensionless, \emph{linearized} Navier-Stokes equation describing the fluid's momentum balance reads
\begin{equation}
\partial_t \vb{u} + 2\,\vu{z} \cross \vb{u} + \partial_t \vb{\Omega} \cross \vb{r} = -\grad p+ E\,\nabla^2 \vb{u},
\label{eq:ns}
\end{equation}
where the unit time is $1/\Omega_0$, the unit length is the semimajor axis $a$ of the spheroidal CMB, $E$ is the Ekman number defined as $E=\nu/\Omega_0 a^2$, and $p$ is the \emph{reduced} pressure related to the physical pressure $P$ through
\begin{equation}
\label{eq:redp}
p = \frac{P}{\rho_f}-\left(\vb{\hat z}\cross\vb{r}\right)\cdot\left(\vb{M}\cross\vb{r}\right)+\Phi,
\end{equation}
where the total angular velocity of the mantle is $\vb{\Omega}=\vb{\hat z}+\vb{M}$, see Eq.~({\ref{eq:Mdef}}) further below, and $\Phi$ is the gravitational potential. The term $\partial_t \vb{\Omega} \cross \vb{r}$ in Eq.~(\mbox{\ref{eq:ns}}) is known as the \emph{Poincar\'e} or \emph{Euler} force, a ficticious force arising due to the mantle's unsteady rotation.
Now, to have a well defined problem we need to determine the mantle's angular velocity $\vb{\Omega}$ by considering the torque balance. The total torque exerted by the fluid core on the mantle is the sum of the pressure (i.e. topographic) and viscous torques (the gravitational torque vanishes since there is no mass redistribution):
\begin{equation}
\vb*{\Gamma} = \int \vb{r} \cross \grad P \dd{V} - \rho_f E \int \vb{r} \cross \nabla^2 \vb{u} \dd{V} \equiv \rho_f \vb*{\gamma},
\end{equation}
where the volume integrals extend over the whole fluid domain and $\vb*{\gamma}$ is dimensionless. The torque $\vb*{\Gamma}$ becomes itself dimensionless once we choose an appropriate unit for mass, which we do in such a way that $\rho_f=1$. The pressure torque involves the physical pressure, however it is desirable to have an expression for the torque that only involves $\vb{u}$. This can be accomplished once the poloidal-toroidal decomposition of the flow is introduced, see Section~\ref{sec:num}. The Ekman number $E$ constitutes one of the control parameters of the system.
\begin{figure}
\centering
\includegraphics[width=0.5\textwidth]{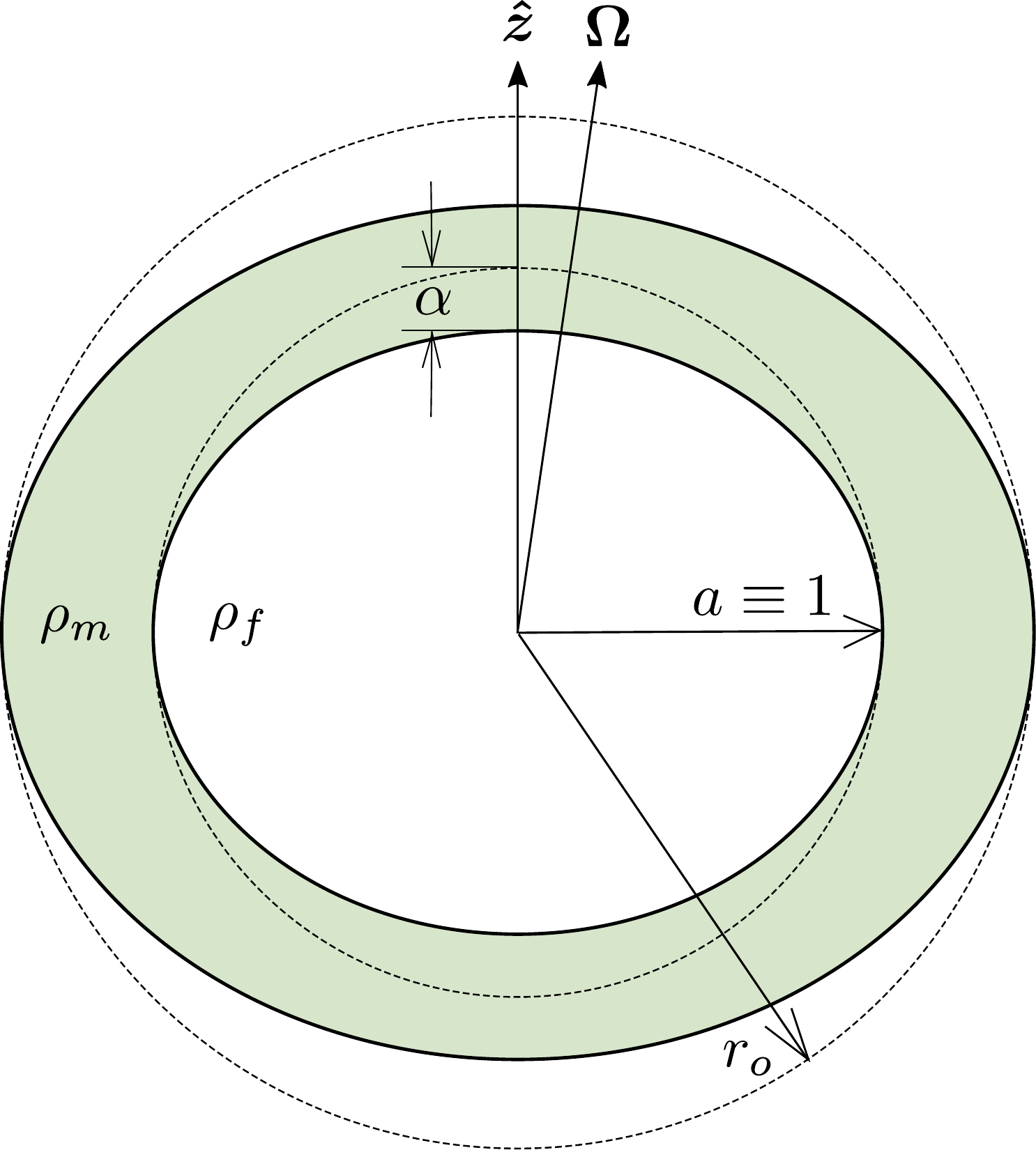}
\caption{Schematic diagram of the two-layer planet model. The fluid core has density $\rho_f$ and the mantle has density $\rho_m$. We choose the vertical $z$ axis in the mantle frame to coincide with the mantle's vertical figure axis. In our choice of units the semimajor axis $a$ of the axisymmetric CMB has unit length. The semimajor axis of the mantle's outer surface, $r_o$, together with the density ratio $\rho_m/\rho_f$ determines the control parameter $q$, see Eq.~\ref{eq:q}.}
\label{fig:schem}
\end{figure}

\subsection{The rigid mantle}
For simplicity we assume that the flattening $\alpha$ (defined as $(a-b)/a$ where $a$ and $b$ are the semimajor and semiminor axes, respectively) of the core-mantle boundary is small and coincides with the flattening of the mantle's outer surface. See Fig.~\ref{fig:schem} for a schematic. If the mantle's moment of inertia around the short axis of symmetry is represented by $\rho_m C$, and the moment of inertia around an equatorial axis by $\rho_m A$ ($\rho_m$ being the mantle's density) then, up to third order in $\alpha$, the \emph{dynamical} flattening is
\begin{equation}
f(\alpha)\equiv\frac{C-A}{A}=\alpha+\frac{1}{2}\,\alpha^2-\frac{1}{4}\,\alpha^3+\order{\alpha^4}.
\end{equation} 
The mantle's spin $\bm \Omega$ is a vector relative to an inertial reference frame, however for our purposes it is convenient to express its components using the mantle's body axes, namely
\begin{equation}
\label{eq:Mdef}
\vb{\Omega} = M_x\,\vu{x}+ M_y\,\vu{y}+ (M_z+1)\,\vu{z}=\vb{M} + \vu{z}
\end{equation}
with $M_x,\,M_y,\,M_z$ being time-dependent and small compared to one. The Liouville equations of motion representing the angular momentum balance in an inertial reference frame reduce to the Euler equations for the motion of the rigid mantle. To first order in $\vb{M}$ they are
\begin{align}
\partial_t{M_x} + f(\alpha)\, M_y & = \frac{\rho_f}{\rho_m A}\gamma_x  \label{eulx}\\
\partial_t{M_y} - f(\alpha)\, M_x & = \frac{\rho_f}{\rho_m A}\gamma_y  \label{euly}\\
\partial_t{M_z} &= \frac{\rho_f}{\rho_m C}\gamma_z.\label{eulz}
\end{align}

The moment of inertia $\rho_m A$ depends on the outer semimajor axis of the mantle $r_0$ and also on the flattening $\alpha$ (the inner semimajor axis of the mantle being unity in our choice of units). To third order in $\alpha$ we can write
\begin{equation}
\frac{\rho_f}{\rho_m A}=\frac{15}{16\pi(r_0^5-1)}\frac{\rho_f}{\rho_m}\left(2+4\alpha+5\alpha^2+5\alpha^3\right),
\end{equation}
consequently, we define a control parameter $q$ as
\begin{equation}
\label{eq:q}
q\equiv\frac{1}{(r_0^5-1)}\frac{\rho_f}{\rho_m},
\end{equation}
which is essentially the inverse of the mantle's mean moment of inertia.
Using recent data for Mercury as an example, we estimate $q\approx0.269$. For Enceladus' icy crust $q\approx 1.6$ and for Earth's mantle $q\approx 0.026$. We have then a total of three control parameters: $E,\,q$ and $\alpha$. Setting $q=0$ will render the torques $\gamma_x,\gamma_y,\gamma_z$ ineffective and as a result the mantle's rotation becomes steady in time. In this case the spectrum of eigenmodes would correspond to the spectrum of inertial modes of a steadily rotating spheroidal cavity. If $\alpha=0$ the boundaries are spherical and therefore only the pressure torque would vanish.

\subsection{Boundary conditions}

We employ a no-slip boundary condition at the CMB and require regularity at the center. Given the near-spherical shape of the CMB it is possible to prescribe the no-slip condition using a Taylor expansion  \citep{rekier2018} as follows:
\begin{equation}
	\label{bc}
	\left.\vb{u}\right|_\textsc{\tiny CMB} = \left.\vb{u}\right|_{r=1} + 
    (r_\textsc{\tiny CMB}-1)\,  \left. \partial_r   \vb{u}\right|_{r=1} +
	{\frac{1}{2}}(r_\textsc{\tiny CMB}-1)^2\,\left. \partial_r^2 \vb{u}\right|_{r=1} +
	{\frac{1}{6}}(r_\textsc{\tiny CMB}-1)^3\,\left. \partial_r^3 \vb{u}\right|_{r=1} = 0,
\end{equation}
where $r_\textsc{\tiny CMB}$ describes the shape of the spheroidal CMB. Up to third order in $\alpha$, it satisfies
\begin{equation}
	\label{eq:cmb1}
	r_\textsc{\tiny CMB}(\theta,\phi)-1 = a_0(\alpha)+a_2(\alpha)\,Y_2^0+a_4(\alpha)\,Y_4^0+a_6(\alpha)\,Y_6^0,
\end{equation}
where $Y_l^m$ are the familiar spherical harmonics and $a_i(\alpha)$ are polynomials of order three in $\alpha$ (see the appendix for their explicit form).
The advantage of this approach is that it is compatible with spherical harmonic expansions for the flow field $\vb{u}$. However, this technique is inadequate if the flattening is comparable or larger than the typical thickness $\sqrt{E}$ of the viscous boundary layer where the components of $\vb{u}$ undergo large spatial variations. In such a situation, the spherical surface $r=1$ would be too far from the Ekman boundary layer for a polynomial series expansion to describe accurately the exponentially decaying flow near the CMB. Therefore we limit our study to combinations of $E$ and $\alpha$ such that $\alpha\ll\sqrt{E}$. This is probably not an issue if stress-free boundary conditions are used since in that case the viscous boundary layers would be mostly suppressed.
We have chosen to extend the Taylor series up to $\order{\alpha^3}$ in order to allow a range as large as possible in the flattening $\alpha$ for a given Ekman number. The resolution of the Taylor expansion is $\sim (\alpha/\sqrt{E})^k$ and should be $\ll 1$ For comparison, a first order calculation ($k=1$) would only allow us to use flattening only up to $\alpha\sim\sqrt{E}/20$, while a third order calculation ($k=3$) allows us to go confidently up to $\alpha\sim\sqrt{E}/5$. A fourth order calculation would only bring a very small increase from this range while greatly increasing the computational cost.

The regularity condition at the origin is better described using the symmetry characteristics of the spherical harmonics, which we introduce below.

\section{Numerical method}
\label{sec:num}

\subsection{Radial and angular discretization}
We deal with oscillatory and possibly damped motions, therefore we write the time dependence as
\begin{equation}
%\vb{u}(\vb{r},t) &=\vb{u}_0(\vb{r})\text{e}^{\lambda t} + \vb{u}_0^\dagger(\vb{r}) \text{e}^{\lambda^\dagger t},\\
%\vb{M}(t) &= \vb{M}_0\text{e}^{\lambda t}+\vb{M}_0^\dagger\text{e}^{\lambda^\dagger t}
\vb{u}(\vb{r},t) =\vb{u}_0(\vb{r})\text{e}^{\lambda t},\,\vb{M}(t) = \vb{M}_0\text{e}^{\lambda t},
\end{equation}
where $\vb{u}_0,\,\vb{M}_0$ and $\lambda=\sigma+i\omega$ are complex valued. We assume an incompressible fluid, hence a divergenceless flow. A poloidal-toroidal decomposition for $\vb{u}_0$ is then adequate:
\begin{equation}
\vb{u}_0(\vb{r}) = \curl{\curl{\mathcal P\vb{r}}}+\curl{\mathcal T\vb{r}}.
\end{equation}
The scalar functions $\mathcal P,\mathcal T$ are in turn expanded into spherical harmonics:
\begin{align}
\label{eq:poltor}
\mathcal P(r,\theta,\phi) &= \sum_{\ell=1}^{\ell_\text{max}} \sum_{m=-\ell}^{\ell}  P_{\ell m}(r)\,Y_\ell^m(\theta,\phi), \\
\mathcal T(r,\theta,\phi) &= \sum_{\ell=1}^{\ell_\text{max}} \sum_{m=-\ell}^{\ell}  T_{\ell m}(r)\,Y_\ell^m(\theta,\phi).
\end{align}

For the radial discretization of the components $P_{\ell m}$ and $T_{\ell m}$ we employ a fast spectral method devised by  \citet{olver2013}.  This method represents the unknown variables using Chebyshev polynomials and Gegenbauer polynomials for their derivatives, its most distinguishing feature is that the resulting matrices are \emph{sparse} as opposed to common spectral collocation methods where the resulting matrices are dense. Explicitly, the Chebyshev expansions for a given spherical harmonic component are 
\begin{align}
P_{\ell m} = \sum_{k=0}^N P_{\ell m}^k\,t_k(r),\\
T_{\ell m} = \sum_{k=0}^N T_{\ell m}^k\,t_k(r),
\end{align}
where $t_k$ is the Chebyshev polynomial of degree $k$. The coefficients $P_{\ell m}^k,\,T_{\ell m}^k$ constitute the unknowns specifying the core flow once the pressure has been eliminated, see section 3.3. The total number of such coefficients is determined by the truncation levels $N$ and $\ell_\text{max}$. 

\subsection{Symmetries}
To comply with the regularity condition at the origin we extend first the radial physical domain from $r\in[0,1]$ to $r\in[-1,1]$ in order to match the natural domain of the Chebyshev polynomials. There is a consideration that must be taken care of, however; for any given radius $r$, $T_{\ell m}(r)$ and $T_{\ell m}(-r)$ are connected by a reflection through the origin and consequently they should possess the same symmetry (parity) as the corresponding spherical harmonic $Y_\ell^m$, i.e.
\begin{equation}
Y_\ell^m(\pi-\theta,\pi+\phi)=(-1)^\ell Y_\ell^m(\theta,\phi).
\end{equation}
Therefore we use only even or odd Chebyshev polynomials depending on whether $\ell$ is even or odd. Obviously the same consideration goes for $P_{\ell m}$.

The problem specified by Eq.~(\ref{eq:ns}) cleanly decouples in the azimuthal wave number $m$ after carrying out the expansions explained above, which is a consequence of the CMB being axisymmetric. This is not the case for different angular degrees $\ell$ that end up coupled by the Coriolis force and by the boundary condition at the CMB.

The solutions for $\vb{u}$ possess a well defined equatorial symmetry: they are either equatorially symmetric, in which case the poloidal scalars $P_{\ell m}$ are such that $\ell-m=$ even while the toroidal scalars $T_{\ell m}$ fulfil $\ell+m=$ odd; or vice versa if the solutions are equatorially antisymmetric. This, together with the symmetry requirements for the Chebyshev polynomials, reduces the size of the problem to one fourth of the original.
In this study we focus mainly with flows associated with nutations, which are equatorially antisymmetric with $m=1$.

\subsection{Pressure torque as a function of flow velocity}
We still need a way to compute the topographic torque as a function of $\vb{u}_0$ in order to solve simultaneously the motion of the fluid core and the rigid mantle. The goal is to find an expression for the spherical harmonic component $\Pi_{\ell m}(\vb{r})$ of the physical pressure involving only $M_0$ as well as the poloidal and toroidal scalars $\mathcal P(\vb{r})$ and $\mathcal T(\vb{r})$, which may contain spherical harmonic components additional to $\ell$ due to the coupling induced by the Coriolis force. This can be accomplished by taking the \emph{consoidal} $\ell, m$ component (i.e. proportional to $\grad Y_l^m$) of Eq.~(\ref{eq:ns}) after expanding all terms in spherical harmonics and then solving for $\Pi_{\ell m}(\vb{r})$. Since these steps are algebraically very cumbersome, we perform them with the help of \classname{TenGSHui}, a symbolic tensor calculus package developed by  \citet{trinh2018}. The resulting expressions are listed in Appendix \ref{sec:apb}.  

\subsection{The generalized eigenvalue problem}
Our unknowns are the set of coefficients $P_{\ell m}^k,\,T_{\ell m}^k$ and the components of $\vb{M}_0$. For the fluid core part we take first the curl of both sides of Eq.~(\ref{eq:ns}), compute the projection along the position vector $\vb{r}$, and multiply by $r^2$ (to avoid singularities at the origin). Reorganizing the terms a bit this is
\begin{equation}
\label{eq:curl1}
r^2\,\vb{r}\cdot \curl{\left( E\nabla^2\vb{u}_0 -2\vu{z}\cross\vb{u}_0   \right)} = \lambda\,r^2\,\vb{r}\cdot \curl{\left(\vb{u}_0+\vb{M}_0\cross\vb{r}\right)}.
\end{equation}
Once the spherical harmonic expansion is carried out the expression above will involve the toroidal coefficients $T_{\ell m}^k$ for the most part plus some poloidal coefficients brought in by the Coriolis force. Taking the curl twice and projecting along $r^4\vb{r}$ brings us to
\begin{equation}
\label{eq:curl2}
r^4\,\vb{r}\cdot\curl\curl\left(E\nabla^2\vb{u}_0 -2\vu{z}\cross\vb{u}_0\right) = \lambda\,r^4\,\vb{r}\cdot\curl\curl\left(\vb{u}_0\right),
\end{equation}
involving mostly the poloidal coefficients $P_{\ell m}^k$ and some toroidal ones. Lastly, we write the angular momentum balance more conveniently as
\begin{align}
M_{+}\left[\lambda^\dagger +i\,f(\alpha)\right] &= \frac{\rho_f}{\rho_m A}\gamma_{+},\\
M_{-}\left[\lambda -i\,f(\alpha)\right] &= \frac{\rho_f}{\rho_m A}\gamma_{-},\label{eq:lio}
\end{align}
where $M_\pm\equiv M_x\pm i\,M_y$, $\gamma_\pm\equiv \gamma_x\pm i\,\gamma_y$, and $(^\dagger)$ denotes the complex conjugate. The reason for this choice is that the quantity $\gamma_+$ depends only on the $m<0$ components of $\vb{u}_0$, therefore if we restrict to $m>0$ we need to consider $M_-$ and $\gamma_-$ only. The torque $\gamma_z$ vanishes as long as $m\ne0$, and since we are focusing on antisymmetric $m=1$ modes, we do not need to consider $M_z$.
The vector of unknowns $\vb{x}$ is then composed with the set of coefficients $\{P_{\ell m}^k\}$, $\{T_{\ell m}^k\}$ together with $M_-$. The problem represented by Eqs.~(\ref{eq:curl1}), (\ref{eq:curl2}) and (\ref{eq:lio}) becomes a generalized eigenvalue problem of the form
\begin{equation}
\label{eq:evp}
\vb{A}\,\vb{x} = \lambda\,\vb{B}\,\vb{x}
\end{equation}
where $\vb{A}$ and $\vb{B}$ are complex square matrices. The boundary condition represented by Eq.~(\ref{bc}) is included by replacing appropiate rows in both $\vb{A}$ and $\vb{B}$, see  \citet{olver2013} for further details. We solve Eq.~(\ref{eq:evp}) numerically using the open-source packages SLEPc  \citep{dalcin2011,slepc-toms,slepc-manual} and MUMPS  \citep{MUMPS01,MUMPS02} employing a \emph{shift and invert} strategy.

\section{Overview of the results}
\label{sec:res}

\subsection{The mode spectrum}
\begin{figure}
\centering
\includegraphics[width=1.15\linewidth]{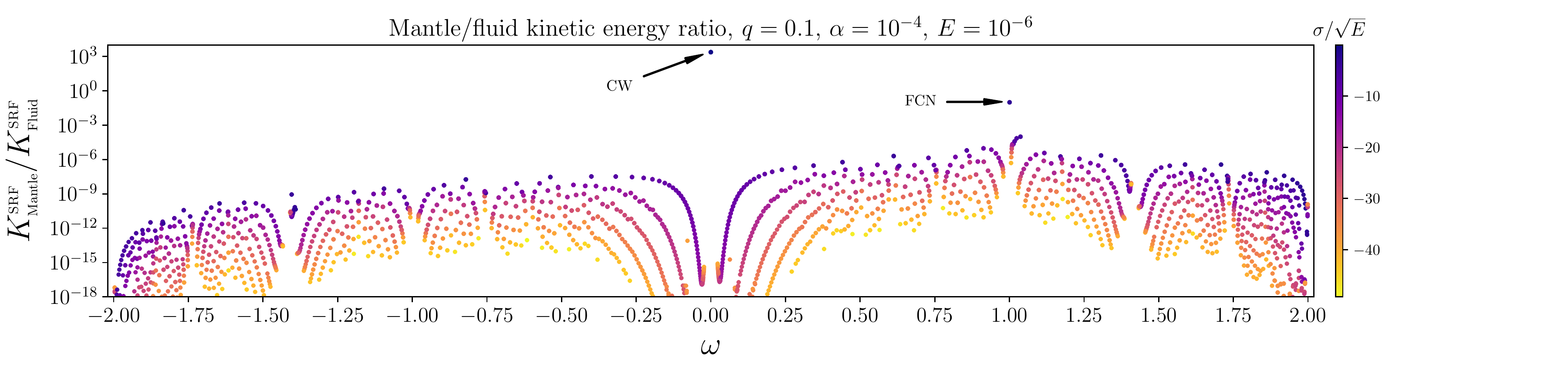}
\caption{Ratio of mantle's kinetic energy to core kinetic energy in the steadily rotating frame (SRF). Horizontal axis is the mode frequency $\omega={\mathcal Im}(\lambda)$, the color scale corresponds to the damping $\sigma={\mathcal Re}(\lambda)$ scaled by $\sqrt{E}$.
The spectrum of our two-layer planet model exhibits naturally inertial modes, including the Free Core Nutation (FCN) and the Chandler Wobble (CW). The latter being the least-damped mode of all.}
\label{fig:first}
\end{figure}

We begin by presenting an overview of the spectrum of the least damped equatorially antisymmetric eigenmodes with $m=1$ when $q=0.1$, $\alpha=10^{-4}$ and $E=10^{-6}$. Figure~\ref{fig:first} shows the kinetic energy of the mantle compared to the total kinetic energy of the fluid core as observed from a reference frame that rotates with the mean rotation rate of the planet, i.e. the steadily rotating frame (SRF).
In Appendix \ref{sec:apc} we provide details on how to transform the flow field as seen from the mantle frame (the frame used in the eigenvalue problem) to the flow field as seen from the SRF.

\noindent Two eigenmodes stand out, the Chandler Wobble (CW) and the Free Core Nutation (FCN). The CW, with a small prograde (i.e. negative in our convention) frequency $\omega$, has most of its kinetic energy as an oscillation of the mantle's spin axis and a small fraction of the energy in the fluid core flow. It is by several order of magnitudes the least damped mode of all and it has no counterpart in the spectrum of purely inertial eigenmodes. The FCN on the other hand, is retrograde with frequency $\omega\sim 1$ and about 10\% of its total kinetic energy is taken by the mantle's motion. This fraction is dependent primarily on $q$ as we discuss further below. In an inertial reference frame the FCN and the CW can be described as mostly a uniformly rotating core flow with an axis slightly different from the mantle's spin axis. The FCN mode resembles the well-known `spin-over' inertial mode with the main difference being that the mantle here is executing an oscillatory motion in addition to steady rotation. The rest of the modes have only a very small fraction of their energy in the mantle's spin oscillation, particularly the prograde ones. All modes, with the exception of the CW, have a counterpart in the spectrum of the purely inertial eigenmodes of a steadily rotating spheroid.

Other features evident in Fig.~\mbox{\ref{fig:first}}, such as the `voids' in the frequency distribution of the eigenmodes (e.g. near $\omega\sim\pm \sqrt{2}$), or their semingly ordered distribution (e.g. around the frequency band $-0.25<\omega<0.25$), are most likely related to the 'quasi-regular' character of some modes. These features, albeit very interesting in themselves, involve many subtleties that are well beyond the scope of our study. Interested readers are encouraged to consult the recent and very detailed discussion by \mbox{\citet{rieutord2018}}.

\begin{figure}
\centering
\subfloat[]{\includegraphics[width=0.5\textwidth]{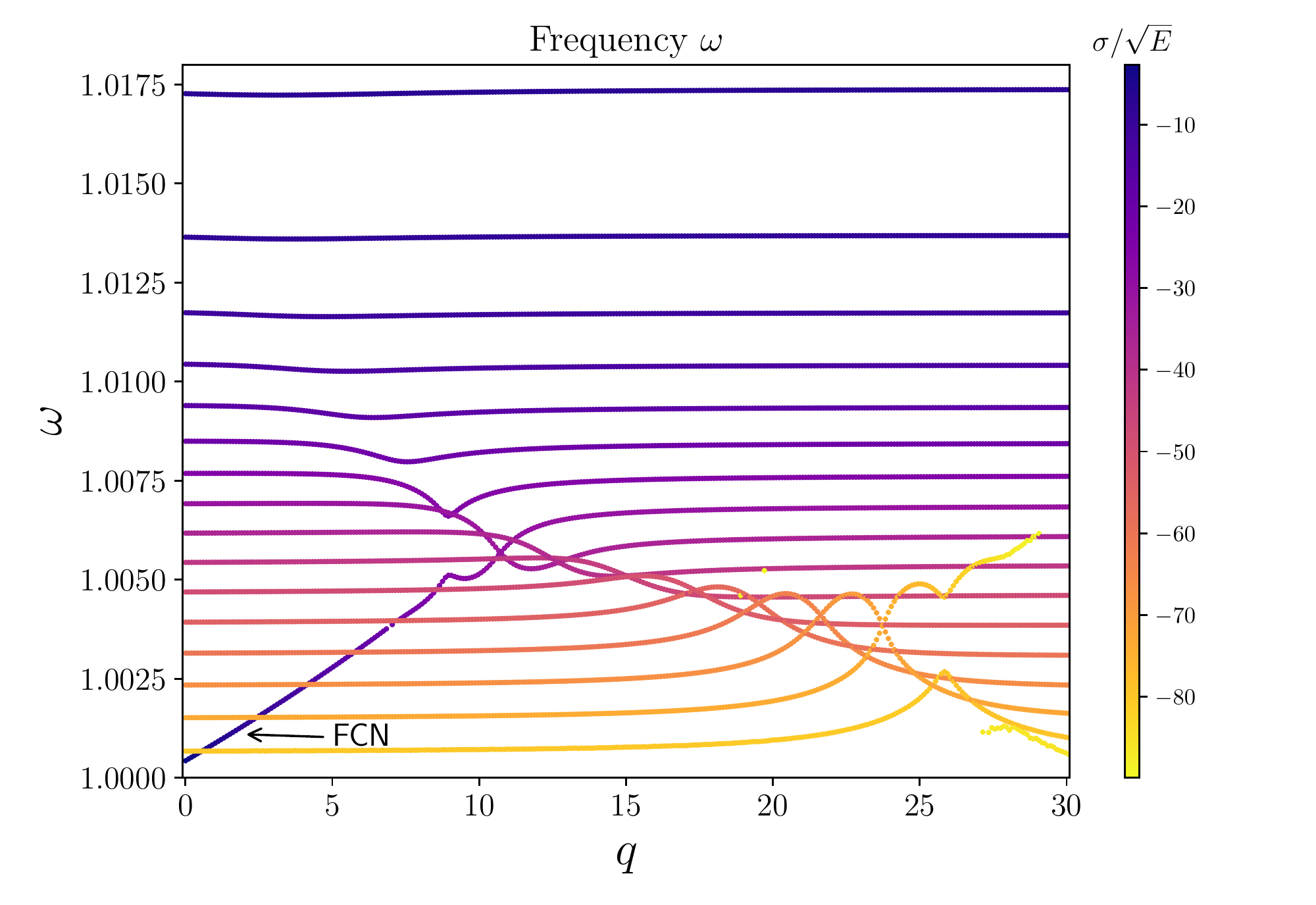}}
\subfloat[]{\includegraphics[width=0.5\textwidth]{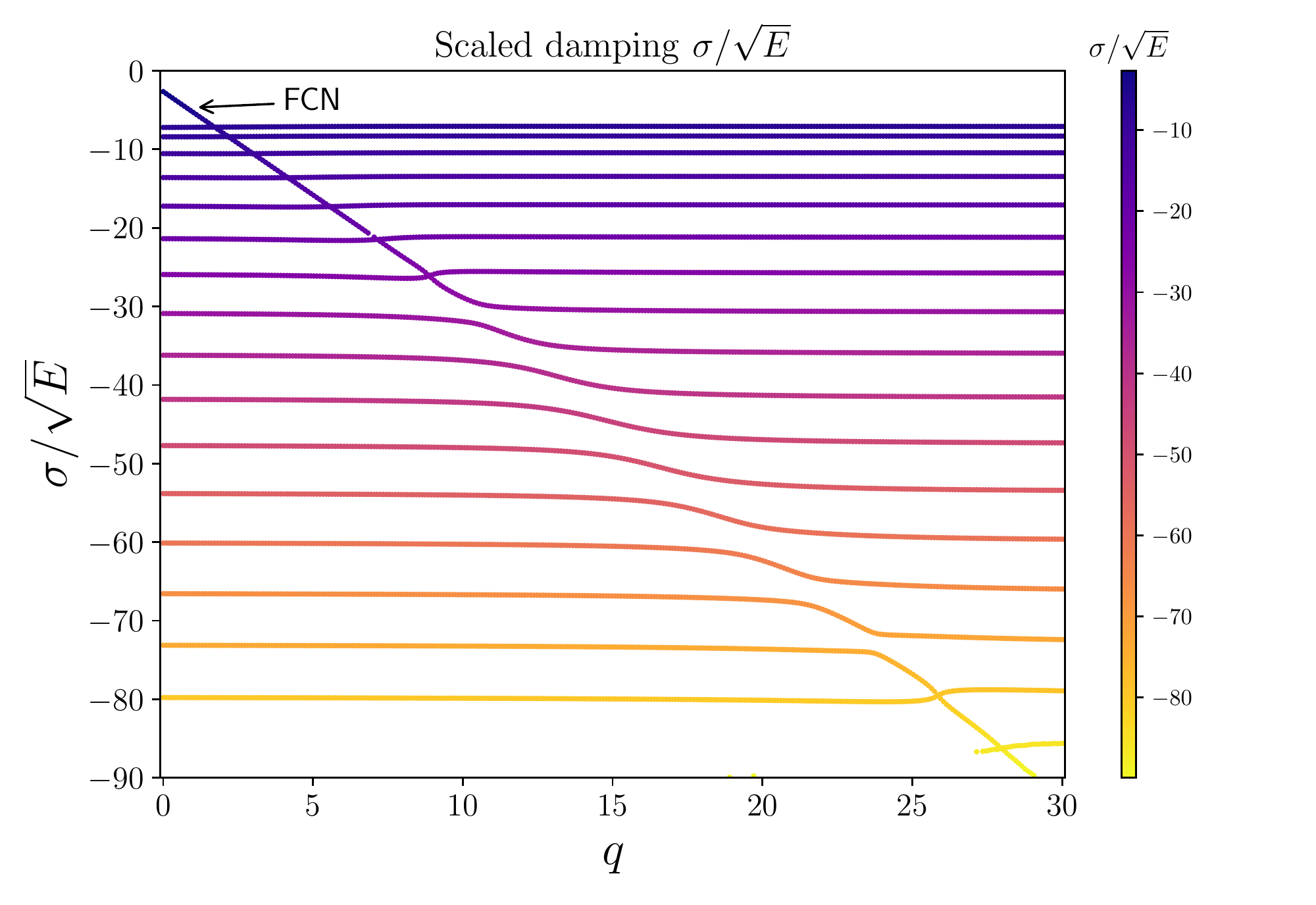}}\\
\subfloat[]{\includegraphics[width=0.5\textwidth]{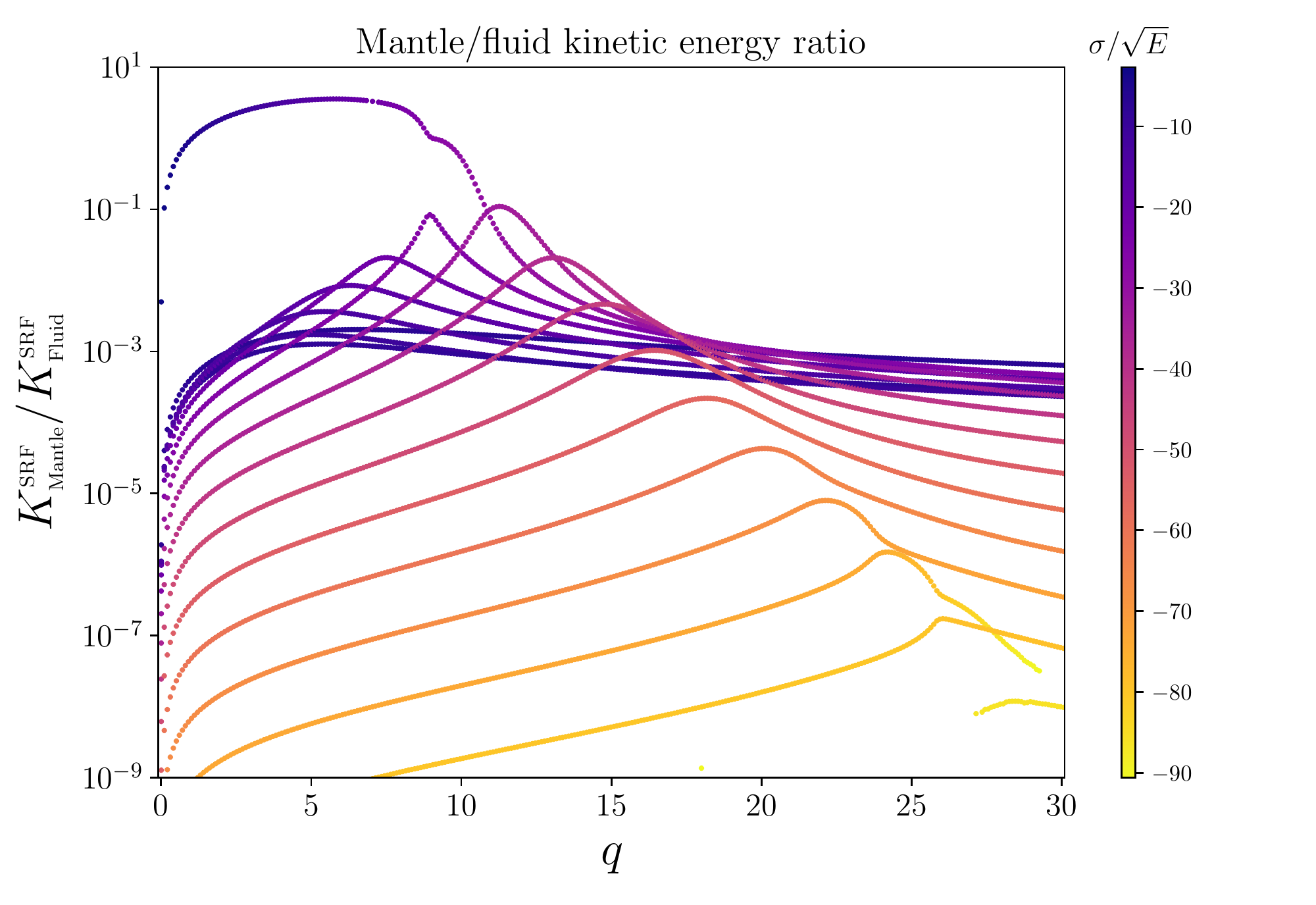}}
\subfloat[]{\includegraphics[width=0.5\textwidth]{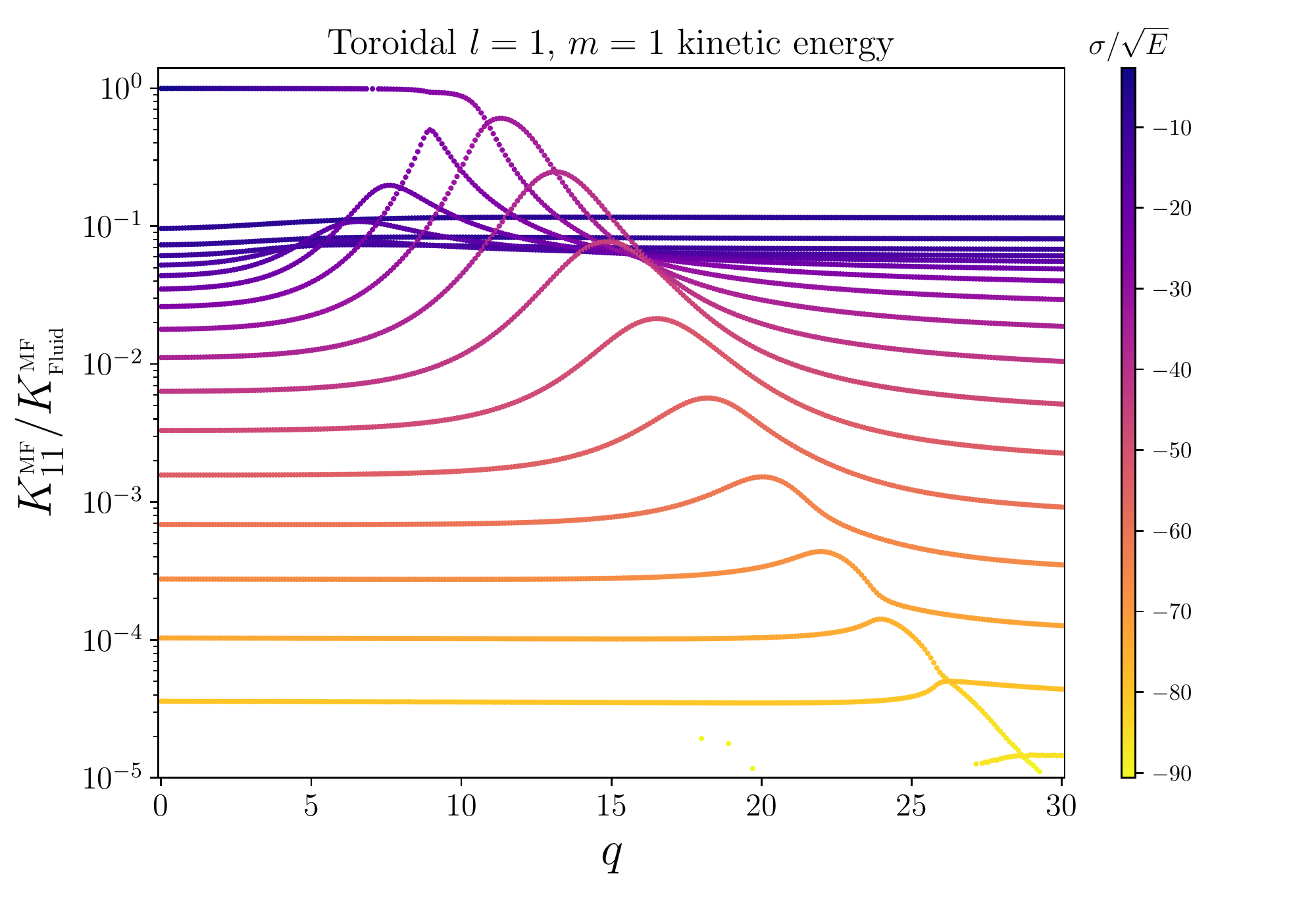}}\\
\caption{As the inverse moment of inertia represented by $q$ is varied, the FCN induces a profound rearrangement of neighbouring modes. The FCN's frequency grows approximately linearly with $q$ when $q\lesssim 1$ as seen in panel (a), while it grows more damped as shown in (b). To ease mode identification, the curves across all panels are colored according to the scaled damping $\sigma/\sqrt{E}$ shown in (b). Here the Ekman number is $E=10^{-6}$ and the flattening is $\alpha=10^{-4}$. See main text for more details.}
\label{fig:main}
\end{figure}

\subsection{Mode interactions}
Now we want to explore the behavior of the modes in the vicinity of the FCN as the control parameter $q$ is varied. We keep the same flattening and Ekman number as before and vary $q$ starting from zero as shown in Fig.~\ref{fig:main}. When $q\sim 0$ the FCN is essentially identical to the spin-over mode and is the second least-damped mode, with the CW being the least-damped one. The toroidal $\ell=1,m=1$ component $T_{11}$ of the core flow is dominant as evidenced by Fig.~\ref{fig:main}(d), which is a typical feature of a flow undergoing mostly a solid body rotation. Both the FCN and the CW exhibit this characteristic. As $q$ continues increasing the FCN's frequency goes up at the same time that it becomes more damped. From Fig.~{\ref{fig:main}}(b) we see that at $q\sim 2$ the FCN is already more damped than the mode at $\omega\approx1.017$, which in turn becomes the second least damped mode after the CW. The fraction of the total energy involved in the FCN's mantle oscillation increases and reaches a maximum around $q\sim 6$ as Fig.~\ref{fig:main}(c) shows.
Interestingly, around $q\sim 8$, the FCN begins interacting with nearby modes and it undergoes `avoided crossings' in frequency while `crossing' in damping, or vice versa. At $q\sim 11$ the toroidal $T_{11}$ component of the FCN is not anymore dominant and therefore the mode cannot be described any longer as a solid body rotation flow. In the range $10<q<30$ other modes undergo similar interactions. In this range frequency crossings take place, while avoided crossings in the damping occur that could rather be described as consecutive `mode bumpings'. The FCN mutates into a mode that is not affected by changes in $q$ any more. Around $q\sim 26$ a mode crossing in damping takes place (near $\sigma/\sqrt{E}\sim -80$) along with a corresponding avoided crossing in frequency. It is possible that similar dynamics can take place again at higher $q$ and more negative damping values.

\begin{figure}
\centering
\includegraphics[width=0.65\textwidth]{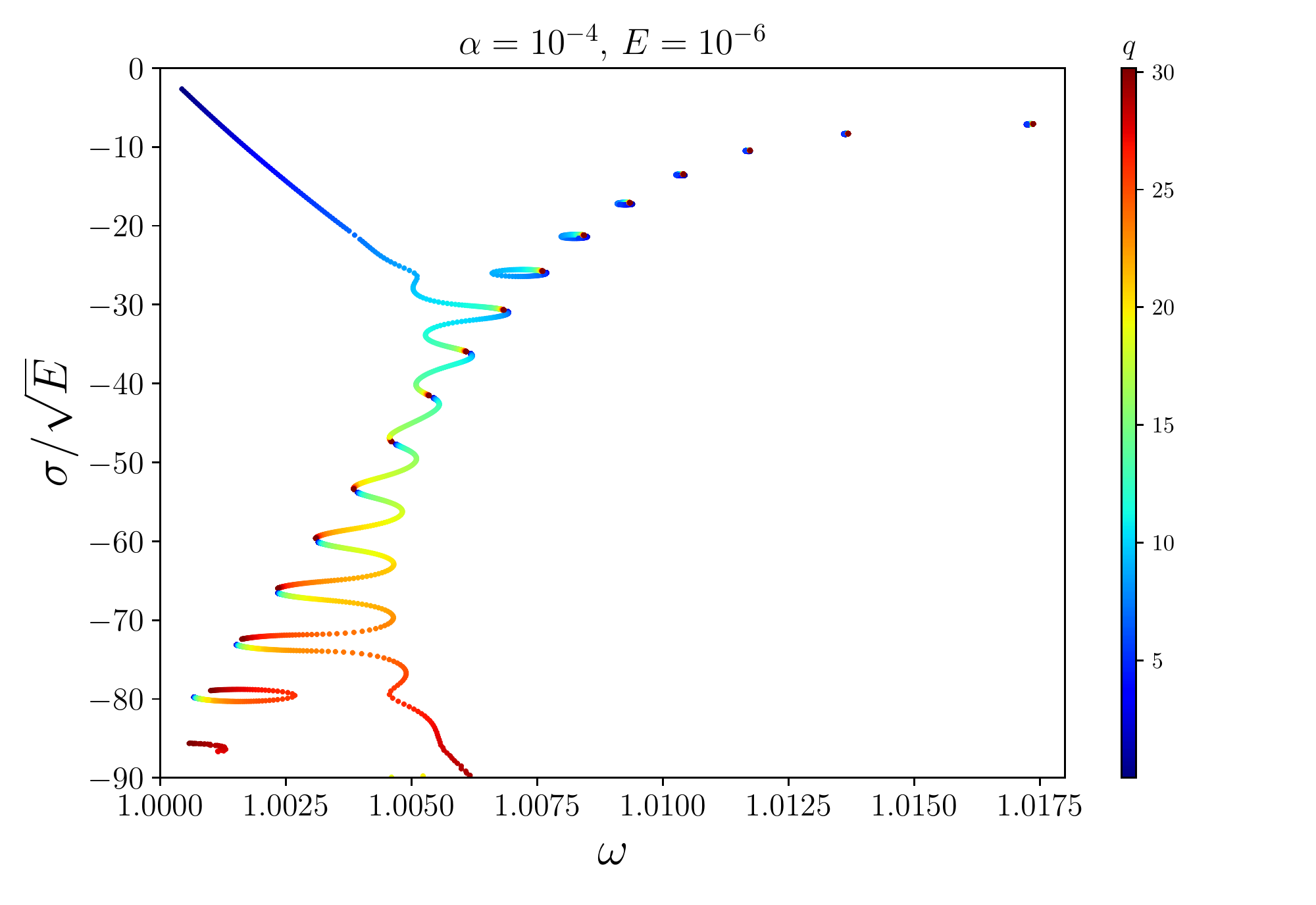}
\caption{Mode trajectories in the $(\omega,\sigma)$ plane as $q$ is varied (indicated by the color scale). It is possible to establish a path from the FCN mode at $q=0$ leading to the most damped mode of Fig.~\ref{fig:main}(b) linking intermediate damping modes and varying $q$ back and forth. Modes performing loops do not lose their identity.}
\label{fig:complex}
\end{figure}

The FCN mode induces a substantial rearrangement of the modes as $q$ is varied. At the same time its own identity is lost in the process and when $q\sim15$ it has taken the place of the mode that initially had $\sigma/\sqrt{E}\sim -30$ when $q=0$. This latter mode in turn morphed into the one that had initially $\sigma/\sqrt{E}\sim -36$ when $q=0$. This exchange appears to continue successively as the modes `bump' each other into more negative damping values. In Fig.~\ref{fig:complex} we show the trajectory that the modes describe in the $(\omega,\sigma)$ plane as $q$ changes. Interestingly, it is indeed possible to follow a path connecting the FCN mode at $q=0$ to the most damped mode visible in Fig.~\ref{fig:main}(b) at $q\sim6$. The modes that form isolated loops in Fig.~\ref{fig:complex} are the ones that do not lose their identity throughout the interactions induced by $q$. We also note that even though the modes at $q=0$ seem to match the `morphed' modes at large $q$, the eigenmodes are not identical since the mantle's kinetic energy is different as Fig.~\ref{fig:main}(c) shows. Only their fluid part match.

\begin{figure}
\centering
\subfloat[]{\includegraphics[width=0.5\textwidth]{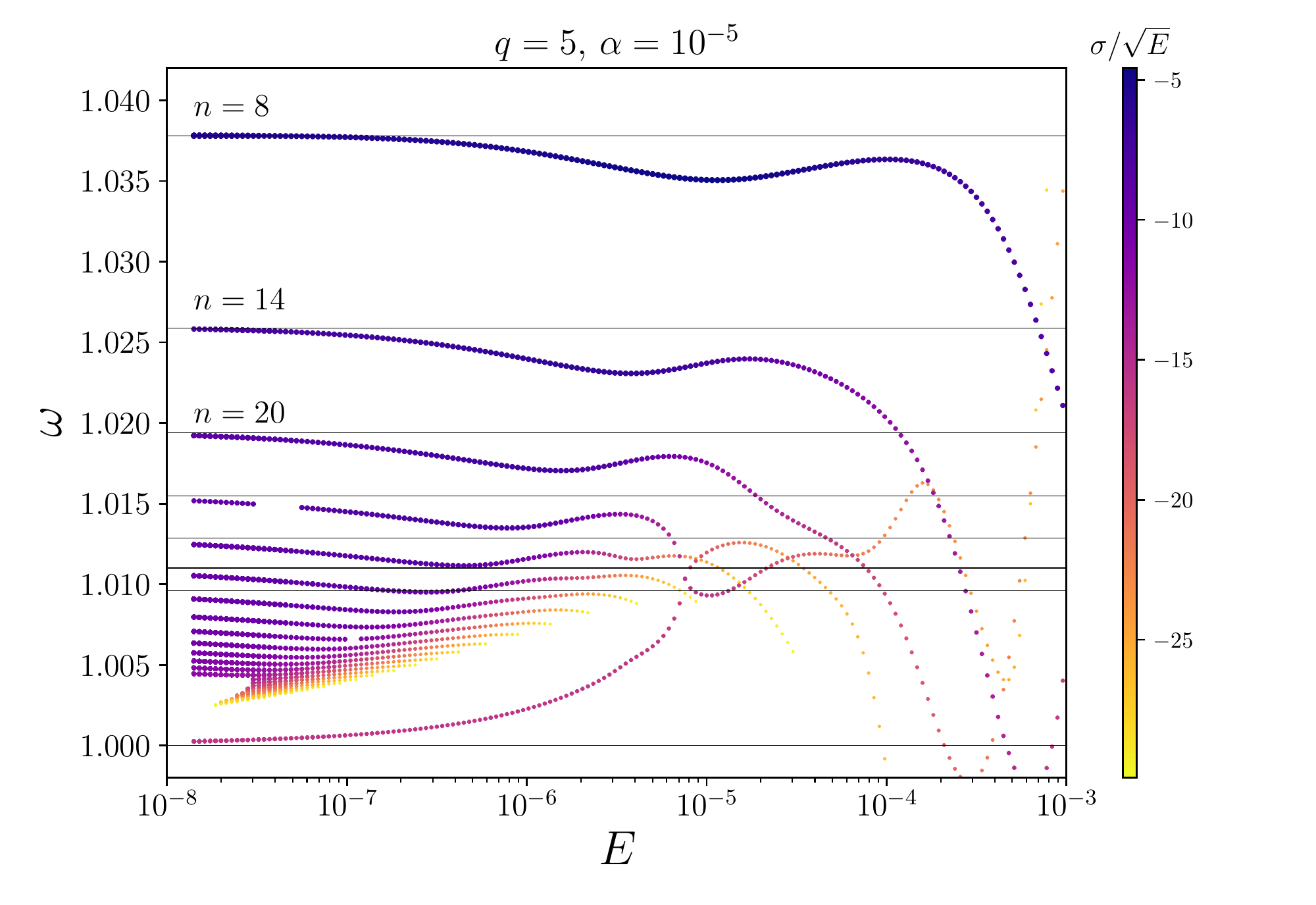}}
\subfloat[]{\includegraphics[width=0.5\textwidth]{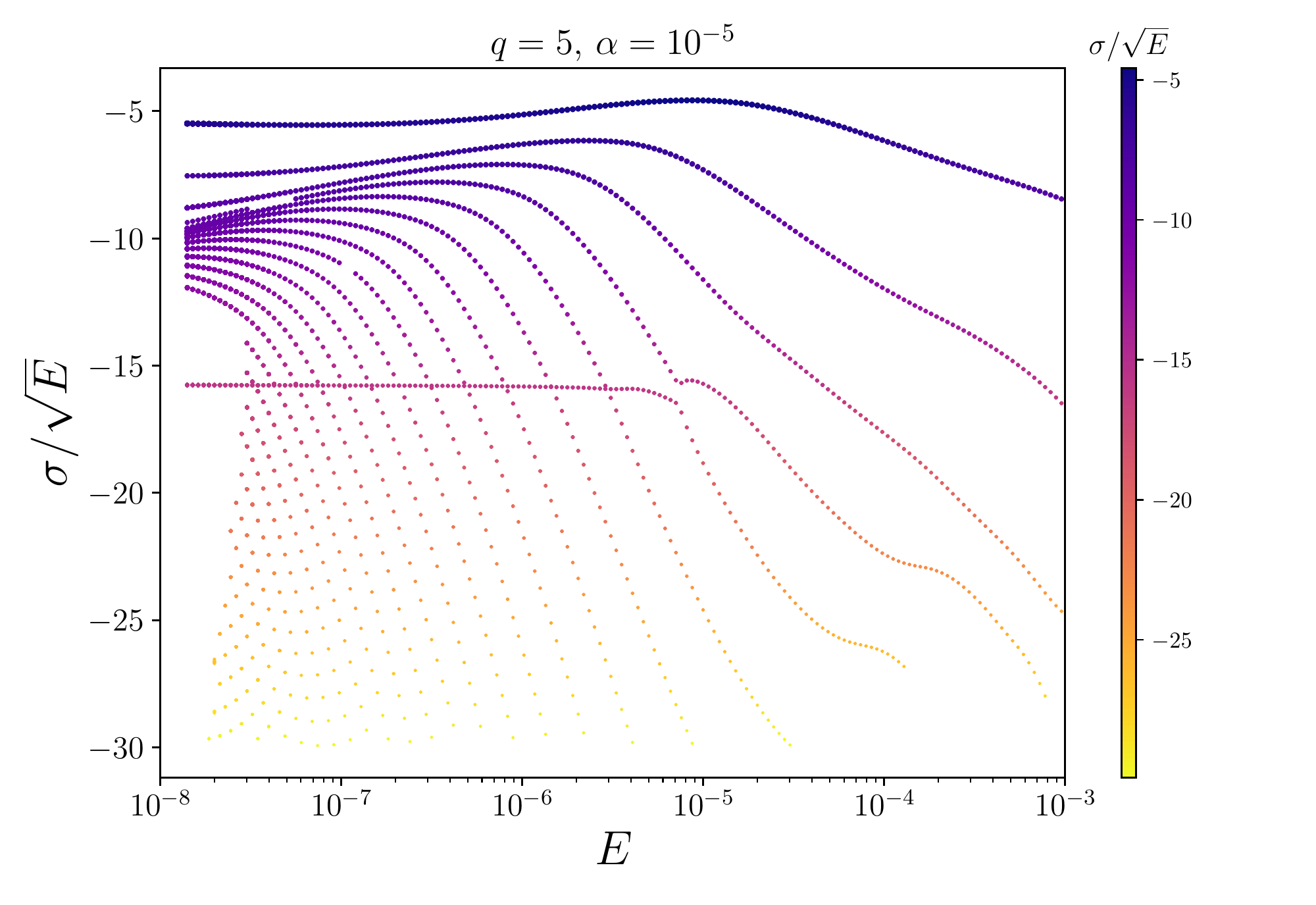}}
\caption{The frequency (a) and scaled damping (b) of eigenmodes near the FCN as the Ekman number is varied. Here $\alpha=10^{-5}$ and $q=5$. Colors indicate the scaled damping $\sigma/\sqrt{E}$. The black horizontal lines in (a) mark the frequency of the theoretical inviscid modes of a sphere for the family of modes with $n=2+6j,\,j=1,2,\dots$ (See main text for further details). The FCN on panel (a) starts near $\omega\sim1.000$ at $E=10^{-8}$ and on panel (b) it starts near $\sigma/\sqrt{E}\sim -16$ at $E=10^{-8}$. As $E$ increases the FCN undergoes a crossing in frequency and an avoided crossing in the damping near $E\sim 10^{-5}$.}
\label{fig:Eramp}
\end{figure}

As already mentioned, the FCN mode at $q=0$ corresponds to the viscous spin-over mode in a steadily rotating spheroid. Apart from the viscous boundary layer and internal shear layers, this mode corresponds to the analytical (inviscid) mode in a full sphere with wavenumbers $(n=2,k=1,m=1)$ using Greenspan's notation  \citep{greenspan1968}. As it turns out, the other modes (at $q=0$) appearing in Fig.~\ref{fig:main}(a) can be linked also to their spherical inviscid counterpart where their $n$ wavenumber differ by \emph{six} between two consecutive modes. This is, they form a family for which $n=6j+2$ with $j=0,1,2,\dots$ and $m=1$ ($m$ being the azimuthal wavenumber) and $k$ is such that the frequency $\omega$ is closest to one, see also Figs.~\ref{fig:Eramp} and \ref{fig:schm}. As $n$ increases, the mode's structure becomes more complex and hence more damped. Interestingly enough, the first three of these modes ($n=8,14,20$) are precisely the ones studied by  \citet{schmitt2006}. According to that study, as the flattening increases the spin-over frequency and the modes in question can undergo frequency crossings, resulting in a similar viscous interaction as in our case.

\subsection{Reducing the Ekman number}
If we reduce the Ekman number, the correspondence explained above can be made more evident since we expect the solutions to approach their inviscid value. We do so in Fig.~\ref{fig:Eramp} where we reduce the Ekman number down to about $E=10^{-7.9}$. However, we need to reduce as well the flattening in order to stay in the range of validity of the spheroidal boundary condition in Eq.~\ref{bc}, therefore we set $\alpha=10^{-5}$ and also set $q=5$. As we discuss later in Section \ref{sec:FCN}, the slope $\partial\omega/\partial q$ of the FCN reduces as $E$ decreases and therefore mode interactions will take place at comparatively higher $q$ values. For instance in Fig.~\ref{fig:Eramp}, where $q=5$, we see already a frequency crossing (with a corresponding avoided crossing in damping) near $E\sim 10^{-5}$. Taking the flattening aside momentarily, this is to be compared with the example at $E=10^{-6}$ in Fig.~\ref{fig:main} where modes begin interacting at $q\sim 8$. Therefore we do not expect mode interactions of this kind to occur at realistic values of $q$ ($q\lesssim 1$) when considering more astrophysically relevant Ekman numbers ($E\lesssim10^{-12}$).
Already at $E=10^{-8}$ the FCN is essentially noninteracting even though $q=5$. In fact, at that Ekman number the frequencies of the $n=2+6j$ inviscid mode family are reassuringly in good agreement with the frequencies of our model as Fig.~\ref{fig:Eramp}(a) shows.

As opposed to the FCN, the CW mode does not appear to interact with any nearby modes in frequency because its damping is very different from its neighbors', which are several orders of magnitude more damped. As we have seen, interactions occur when modes get close enough to each other in the $(\omega,\sigma)$ plane. In fact, although the CW's frequency does change with $q$ and crosses other modes' frequencies, its damping remains small even for moderate $q$, so no interaction occurs. We present a more detailed study of the CW's frequency and damping as functions of $q,\,\alpha$ and $E$ in Section \ref{sec:CW}.

\section{The Free Core Nutation (FCN) mode}
\label{sec:FCN}

We turn now to the dependence of the FCN mode on the parameters $q,\,\alpha$ and $E$ away from the interaction region, which is more relevant for planets. The frequency $\omega_\textsc{\tiny fcn}$ and damping $\sigma_\textsc{\tiny fcn}$ are approximately linear with $q$ as long as there are no other modes with similar $(\omega,\sigma)$ that could interact with the FCN. This translates to $q\lesssim 0.1$ for Ekman numbers in the range $10^{-8}<E<10^{-4}$, and flattenings in the range $10^{-7}<\alpha<10^{-4.6}$. Within those limits we perform linear fits against $q$ following:
\begin{equation}
\begin{split}
\omega_\textsc{\tiny fcn}-\omega_* &= b_\omega (\alpha,E) + q\,m_\omega(\alpha,E),\\
\sigma_\textsc{\tiny fcn} &= b_\sigma (\alpha,E)+ q\,m_\sigma(\alpha,E),
\end{split}
\label{eq:fcn1}
\end{equation}
where $\omega_*$ is the theoretical inviscid spin-over mode frequency in a spheroid with flattening $\alpha$ (computed to third order in $\alpha$). The intercepts $b_\omega$ and $b_\sigma$ represent the frequency and the damping of the viscous spin-over mode, therefore suitable for direct comparison with the viscous corrections computed by  \citet{zhang2004}.

\begin{figure}
\centering
\subfloat[]{\includegraphics[width=0.5\textwidth]{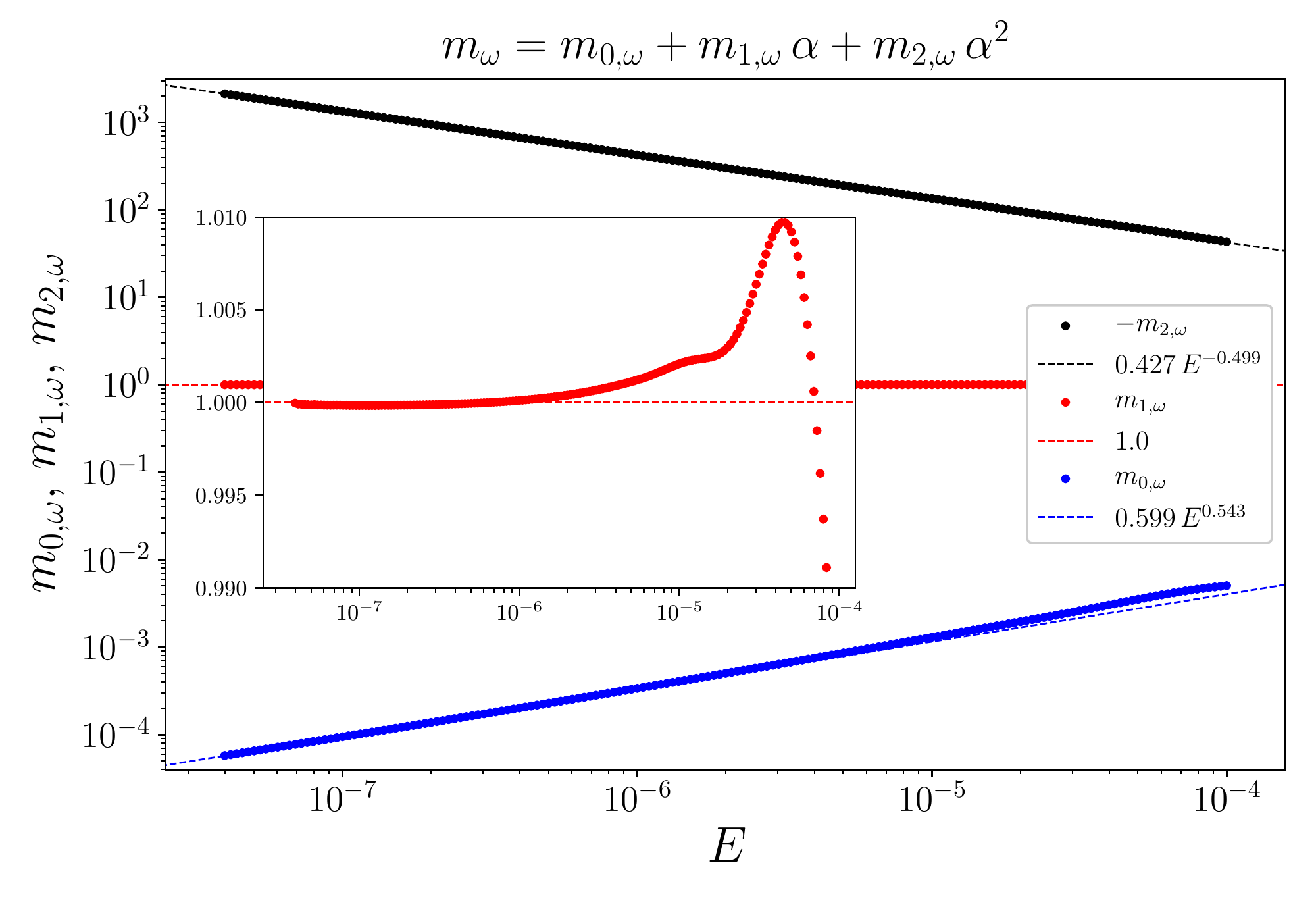}}
\subfloat[]{\includegraphics[width=0.5\textwidth]{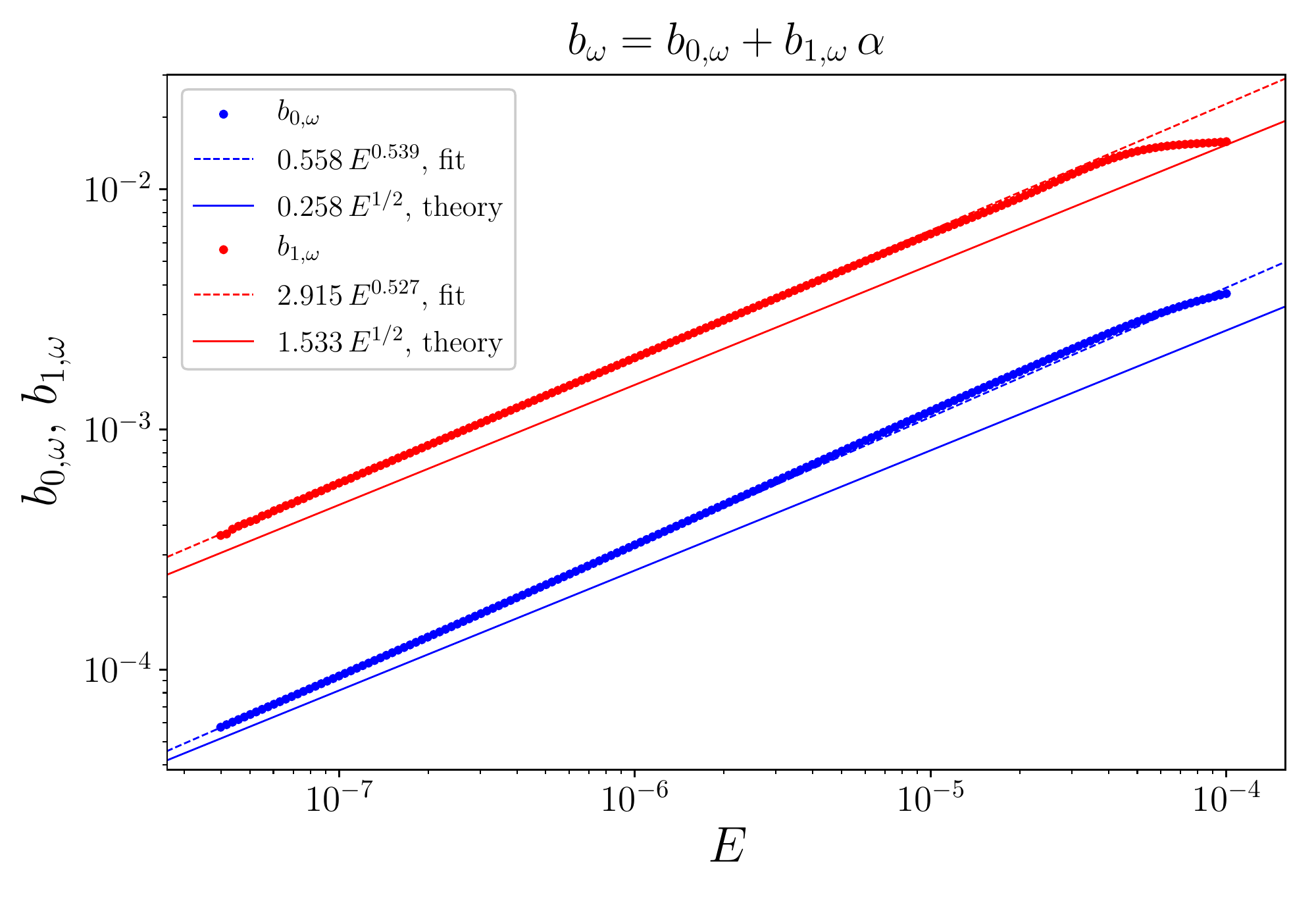}}\\
\caption{Fit coefficients for $m_\omega$ (a) and $b_\omega$ (b) corresponding to the FCN mode frequency $\omega_\textsc{\tiny fcn}$, see Eqs.(\ref{eq:fcn1}) and (\ref{eq:fcn2}). Dots represent the numerical results (after performing linear fits against $q$) and dashed lines are the power law fits against $E$. Inset panel in (a) is a zoom to reveal more detail on $m_{1,\omega}$. The continuous lines in panel (b) represent the theoretical viscous corrections of a steadily rotating spheroid as computed by  \citet{zhang2004}.}
\label{fig:omega_fcn}
\end{figure}

\begin{figure}
\centering
\subfloat[]{\includegraphics[width=0.5\textwidth]{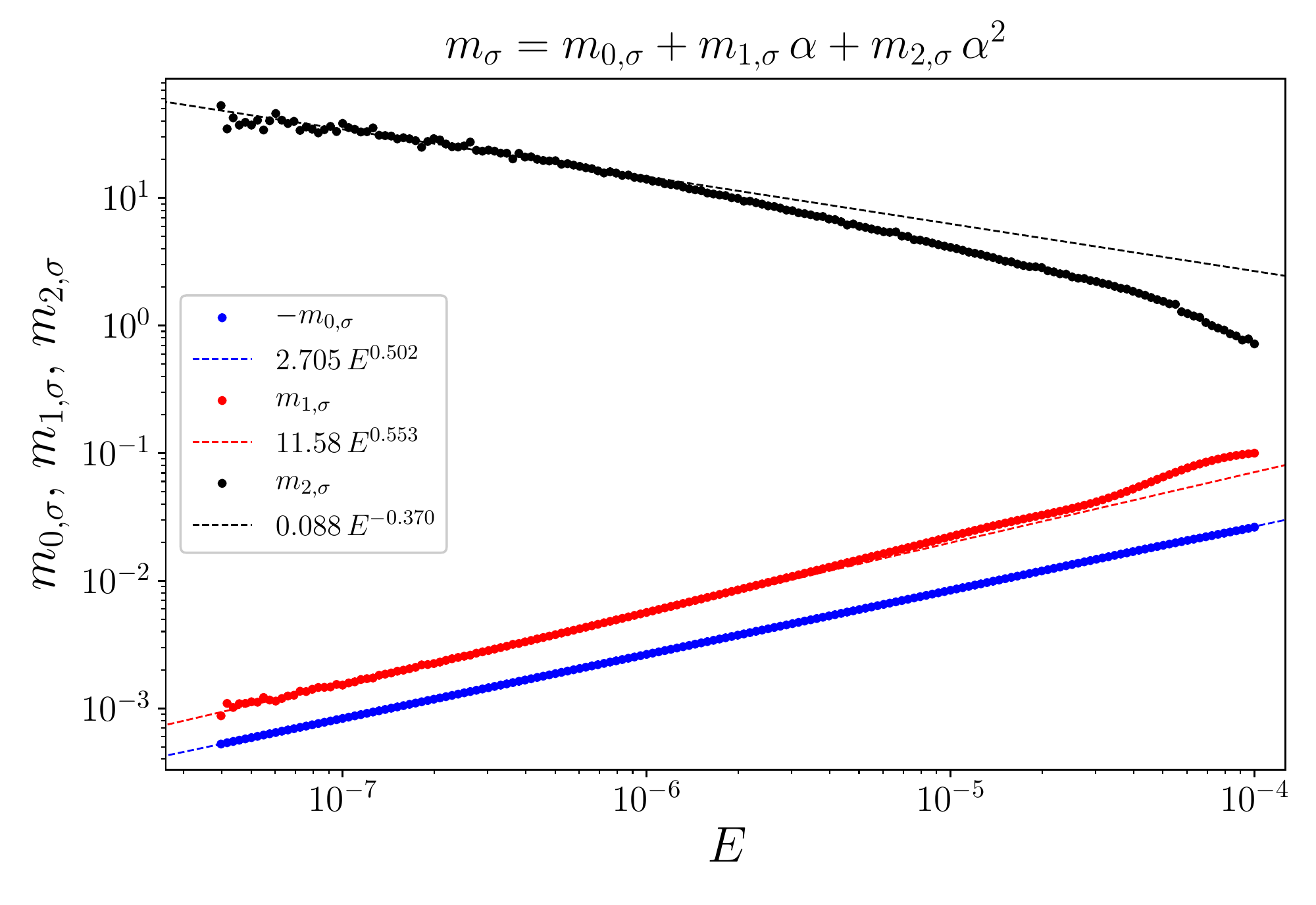}}
\subfloat[]{\includegraphics[width=0.5\textwidth]{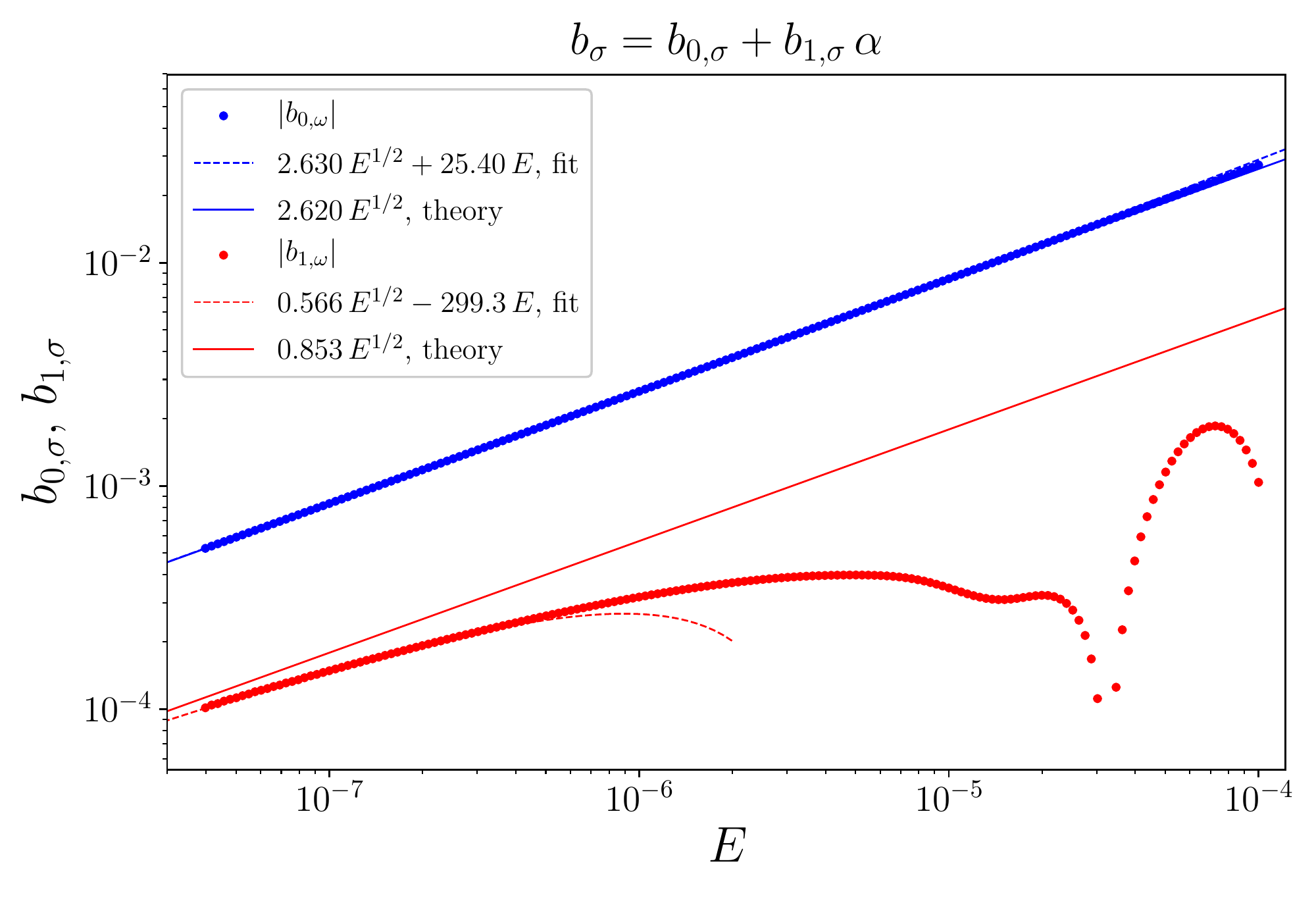}}\\
\caption{Fit coefficients for $m_\sigma$ in panel (a) and $b_\sigma$ in panel (b) corresponding to the FCN mode damping $\sigma_\textsc{\tiny fcn}$, see Eqs.(\ref{eq:fcn1}) and (\ref{eq:fcn3}). Dots represent the numerical results (after performing linear fits against $q$) and dashed lines are the power law fits against $E$. In panel (b) the continuous lines represent the expectations from  \citet{zhang2004}, only with the sign changed. Our results for $b_{0,\sigma}$ are in excellent agreement with the theoretical expectation while $b_{1,\sigma}$ only begins approaching the theoretical scaling towards the lower Ekman numbers we explored.}
\label{fig:sigma_fcn}
\end{figure}

In turn, the slopes and intercepts $m_\omega, b_\omega$ and $m_\sigma, b_\sigma$ appear to be either linear or quadratic in $\alpha$ if $E$ is fixed,
 so we perform the following additional fits:
\begin{equation}
\begin{split}
m_\omega(\alpha,E) &= m_{0,\omega}(E) + \alpha\,m_{1,\omega}(E) + \alpha^2\,m_{2,\omega}(E),\\
b_\omega(\alpha,E) &= b_{0,\omega}(E) + \alpha\,b_{1,\omega}(E),
\end{split}
\label{eq:fcn2}
\end{equation}
together with
\begin{equation}
\begin{split}
m_\sigma(\alpha,E) &= m_{0,\sigma}(E) + \alpha\,m_{1,\sigma}(E) + \alpha^2\,m_{2,\sigma}(E),\\
b_\sigma(\alpha,E) &= b_{0,\sigma}(E) + \alpha\,b_{1,\sigma}(E).
\end{split}
\label{eq:fcn3}
\end{equation}
Lastly, we establish approximate scalings of the coefficients $m_{i,\omega},b_{i,\omega}$ and $m_{i,\sigma},b_{i,\sigma}$ $(i=0,1,2)$ with
respect to the Ekman number $E$. The resulting scaling laws for $\omega_\textsc{\tiny fcn}$ and $\sigma_\textsc{\tiny fcn}$ are shown
in Fig.~\ref{fig:omega_fcn} and Fig.~\ref{fig:sigma_fcn}, respectively.

A few comments are in order. First, from Fig.~\ref{fig:omega_fcn}(a), we note that $m_{0,\omega}(E)$ follows approximately a power law once $E$ is below $\sim10^{-6}$, therefore we perform a power law fit (quoted in the figure) for $E$ in the range $10^{-8}<E<10^{-6}$. In Fig,~\ref{fig:omega_fcn}(a), $m_{1,\omega}$ seems to level off at a value very close to unity, which is expected as we discuss in the appendix. The behaviour of $m_{2,\omega}$ is rather surprising since it scales with a negative power of the Ekman number, which means that the magnitude of $m_{2,\omega}$ \emph{increases} as the $E$ decreases, while we expected $m_{2,\omega}$ to at least level off to a constant value near $-1/2$ for asymptotically small $E$. This scaling is likely valid in only the limited range we used for the flattening and the Ekman number as we discuss further below. The power law fits for $b_\omega$ shown in Fig.~\ref{fig:omega_fcn}(b) are in reasonable agreement with the asymptotically vanishing viscosity correction of the spin-over frequency computed by  \citet{zhang2004}. Summarizing, we approximate the FCN's frequency through $m_\omega$ and $b_\omega$ as
\begin{equation}
\begin{split}
\omega_\textsc{\tiny FCN} &=m_\omega\,q + b_\omega,\,\textrm{where}\\
m_\omega &\approx 0.599\,E^{0.543}+\alpha-0.427\,\alpha^2\,E^{-0.499},\,\textrm{and}\\
b_\omega &\approx 0.558\,E^{0.539}+2.915\,\alpha\,E^{0.527}.
\label{eq:omega_fcn}
\end{split}
\end{equation}

We move on now to the linear fits of the damping $\sigma_\textsc{\tiny FCN}$ against $q$. We show the resulting slopes $m_\sigma$ and intercepts $b_\sigma$ in Fig.~\ref{fig:sigma_fcn}. As there is no prediction of what the damping of the FCN mode might be, theoretically or otherwise, we can only compare the intercepts $b_\sigma$ as they should reduce to the damping of the spin-over mode in a spheroid.  Regarding $m_\sigma$, in Fig.~\ref{fig:sigma_fcn}(a), we find a similar situation as before, the $m_{2,\sigma}$ coefficient seems to follow an inverse power law scaling with $E$, albeit with a smaller magnitude exponent and amplitude compared to $m_{2,\omega}$. Below $E\sim10^{-7}$ we start seeing the effects of numeric round-off errors, which affect more high-order quantities such as $m_{2,\sigma}$ compared to $m_{0,\sigma}$, for example. For this reason we perform fits for $m_{1,\sigma}$ and $m_{2,\sigma}$ in the range $10^{-7.5}<E<10^{-6}$ only. In Fig.~\ref{fig:sigma_fcn}(b) the fit for $b_{0,\sigma}$ matches the theoretical asymptotic expectation rather satisfactorily, however the observed power law for $b_{1,\sigma}$ only begins to approach the asymptotic expectation at the lowest end of our Ekman number range. At the higher end, above $E\sim10^{-4.5}$, our numerical result for $b_{1,\sigma}$ has the opposite sign compared to expectation, which seems to be also the case in the study by  \citet{schmitt2006}, see their Figure 10(a). In sum, we find that the damping of the FCN is approximately
\begin{equation}
\begin{split}
\sigma_\textsc{\tiny FCN} &=  m_\sigma\,q + b_\sigma,\,\textrm{where}\\
m_\sigma &\approx -2.705\,E^{0.502}+11.58\,\alpha\,E^{0.553} +0.088\,\alpha^2\,E^{-0.37},\,\textrm{and}\\
b_\sigma &\approx -2.630\,E^{1/2}-25.40\,E + \alpha\,\left(-0.566\,E^{1/2}+299.3\,E\right).
\label{eq:sigma_fcn}
\end{split}
\end{equation}

\section{The Chandler Wobble (CW) mode}
\label{sec:CW}
\begin{figure}
\centering
\subfloat[]{\includegraphics[width=0.5\textwidth]{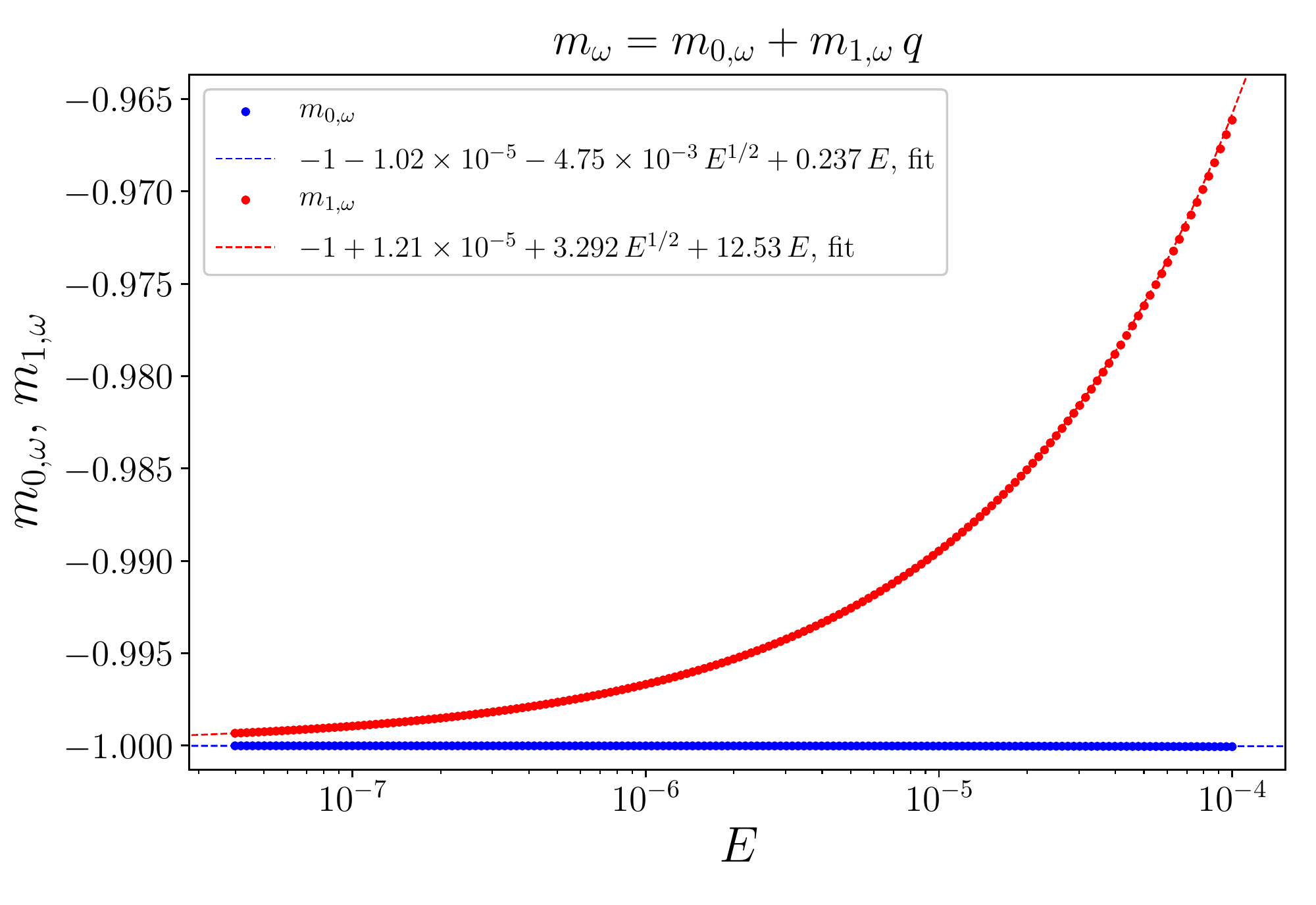}}
\subfloat[]{\includegraphics[width=0.5\textwidth]{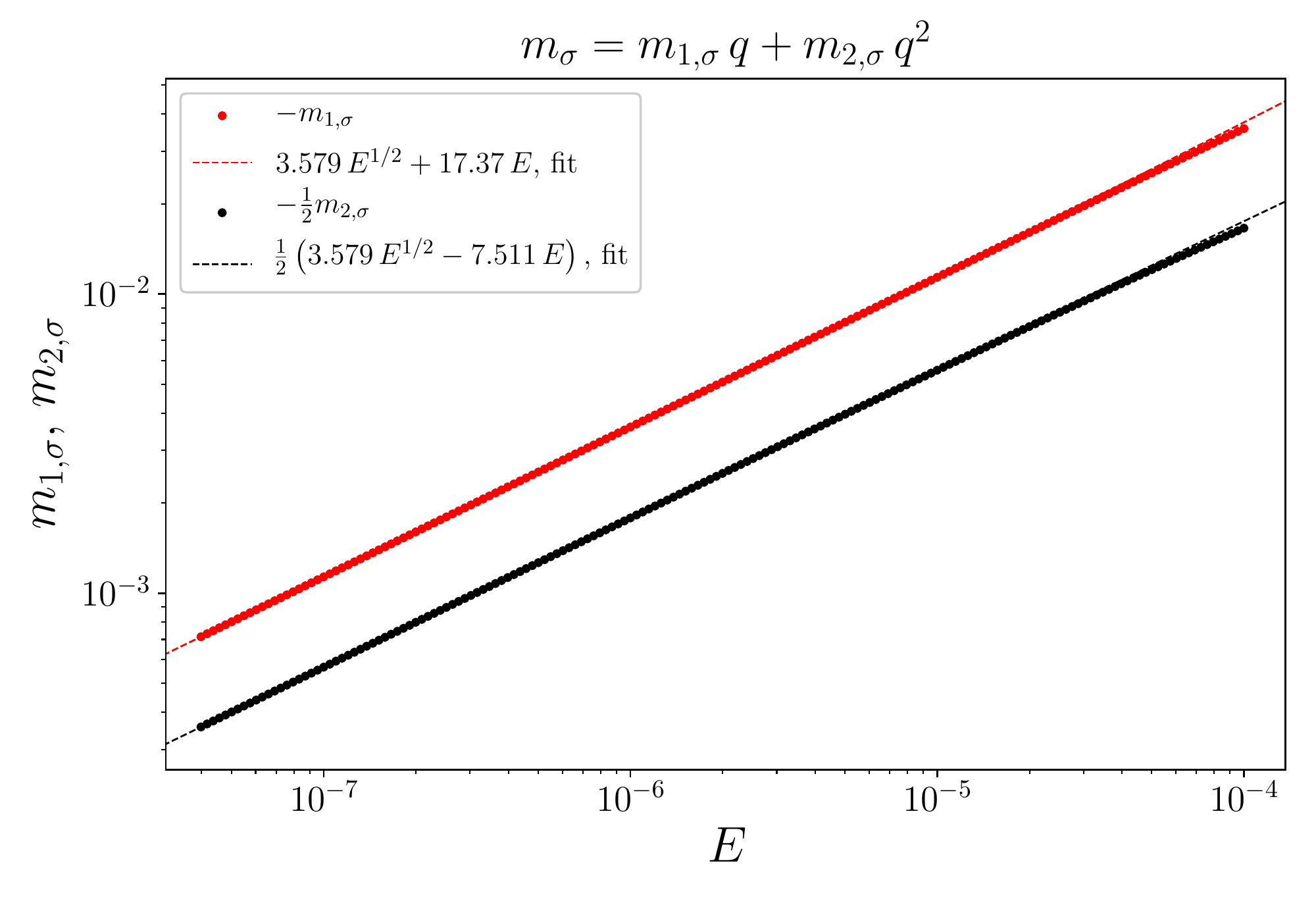}}\\
\caption{Fits for $m_\omega$ (a) and $m_\sigma$ (b) corresponding to the Chandler Wobble mode frequency $\omega_\textsc{\tiny CW}$ and damping $\sigma_\textsc{\tiny CW}$, respectively. See Eqs.(\ref{eq:cw1}) and (\ref{eq:cw2}). In panel (b) $m_{2,\sigma}$ and its corresponding fit have been deliberately multiplied by $1/2$ in order to distinguish it visually from $m_{1,\sigma}$.  }
\label{fig:m_cw}
\end{figure}

There appears to be no mode interactions between the CW and other modes as $q$ is varied. In fact, there are other modes with similarly low frequencies as the CW, but the only ones with potentially low damping that could interact happen to be equatorially symmetric, as can be seen in Fig.~\ref{fig:schm}(a), hence completely decoupled from the CW. The frequency and damping of the CW appear to be directly proportional to the flattening $\alpha$ if $q$ and $E$ are fixed, at least for the parameter range we tested, so it is convenient to perform linear fits against $\alpha$ first:
\begin{equation}
\begin{split}
\omega_\textsc{\tiny cw} &= m_\omega (q,E)\,\alpha,\\
\sigma_\textsc{\tiny cw} &= m_\sigma (q,E)\,\alpha.
\end{split}
\label{eq:cw1}
\end{equation}
The proportionality constants $m_\omega$ and $m_\sigma$ themselves turn out to be approximately linear or quadratic in $q$: 
\begin{equation}
\begin{split}
m_\omega(q,E) &= m_{0,\omega}(E) + q\,m_{1,\omega}(E),\\
m_\sigma(q,E) &= q\,m_{1,\sigma}(E) + q^2\,m_{2,\sigma}(E).
\end{split}
\label{eq:cw2}
\end{equation}
Here we do analogously to the FCN case by establishing the approximate scalings of the coefficients $m_{i,\omega}$ and $m_{i,\sigma}$ $(i=0,1,2)$ appearing in Eq.~(\ref{eq:cw2}) with respect to the Ekman number $E$. Figure \ref{fig:m_cw} shows the results.

As the Ekman number decreases the CW frequency approaches the theoretical inviscid value, i.e., the proportionality constants $m_{0,\omega}$ and $m_{1,\omega}$ get both close to $-1$ following approximately a familiar $E^{1/2}$ scaling. The same scaling applies to the damping as Figure~\ref{fig:m_cw}(b) shows.
Summarizing, we find that the Chandler Wobble frequency is
\begin{equation}
\begin{split}
\omega_\textsc{\tiny CW}=\alpha &\left[-1-1.02\cdot10^{-5}-4.75\cdot10^{-3}\,E^{1/2}+0.237\,E \right. \\
                                      &\,\, \left. +q\left(-1+1.21\cdot10^{-5}+3.292\,E^{1/2}+12.53\,E \right) \right],
\label{eq:omega_cw}
\end{split}
\end{equation}
and its damping is
\begin{equation}
\sigma_\textsc{\tiny CW}=\alpha\,q\,\left[-3.579\,E^{1/2}-17.37\,E+q\left(-3.579\,E^{1/2}+7.511\,E \right) \right].
\label{eq:sigma_cw}
\end{equation}

\section{Discussion}
\label{sec:disc}

\begin{figure}
\centering
\subfloat[]{\includegraphics[width=0.5\textwidth]{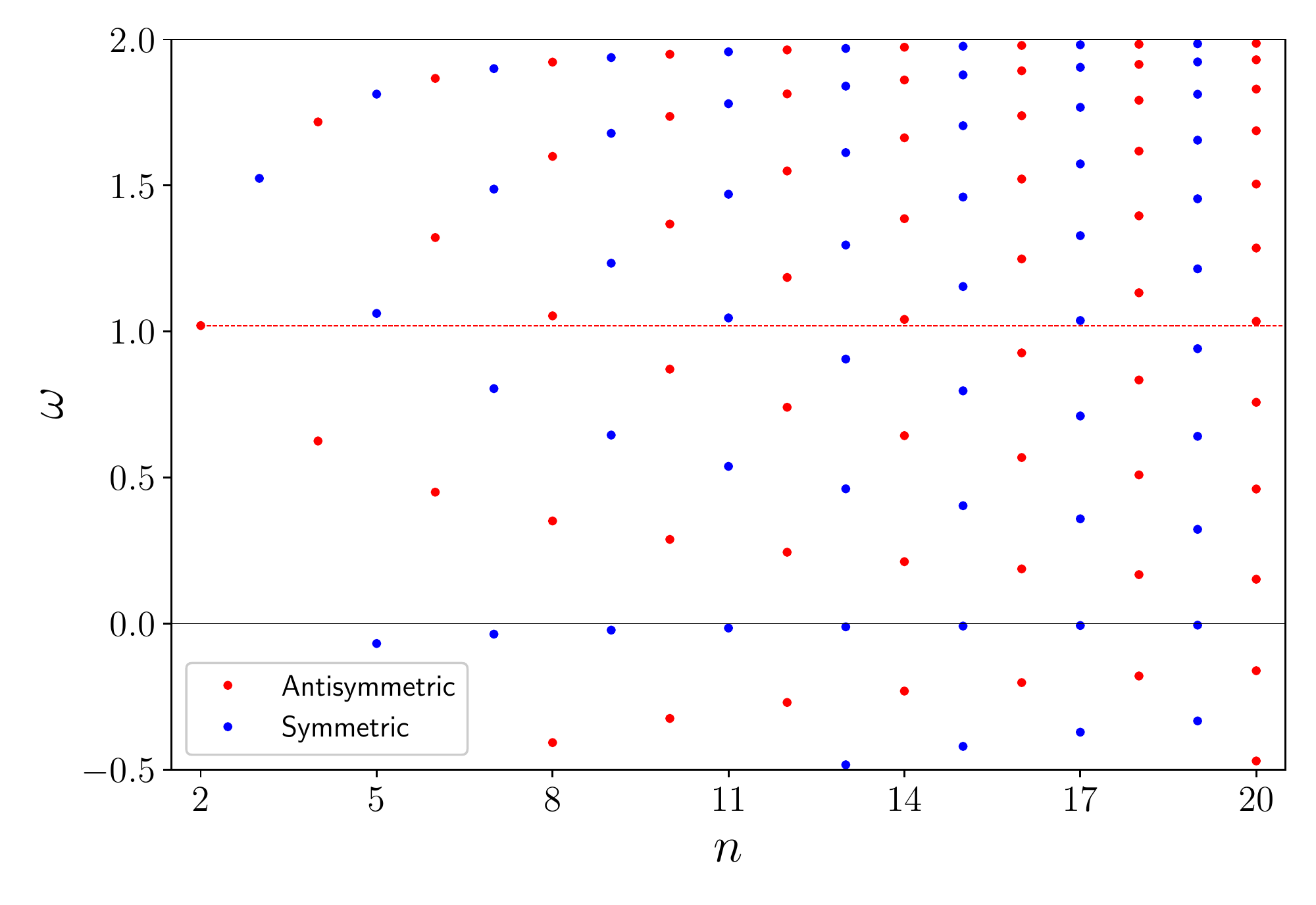}}
\subfloat[]{\includegraphics[width=0.5\textwidth]{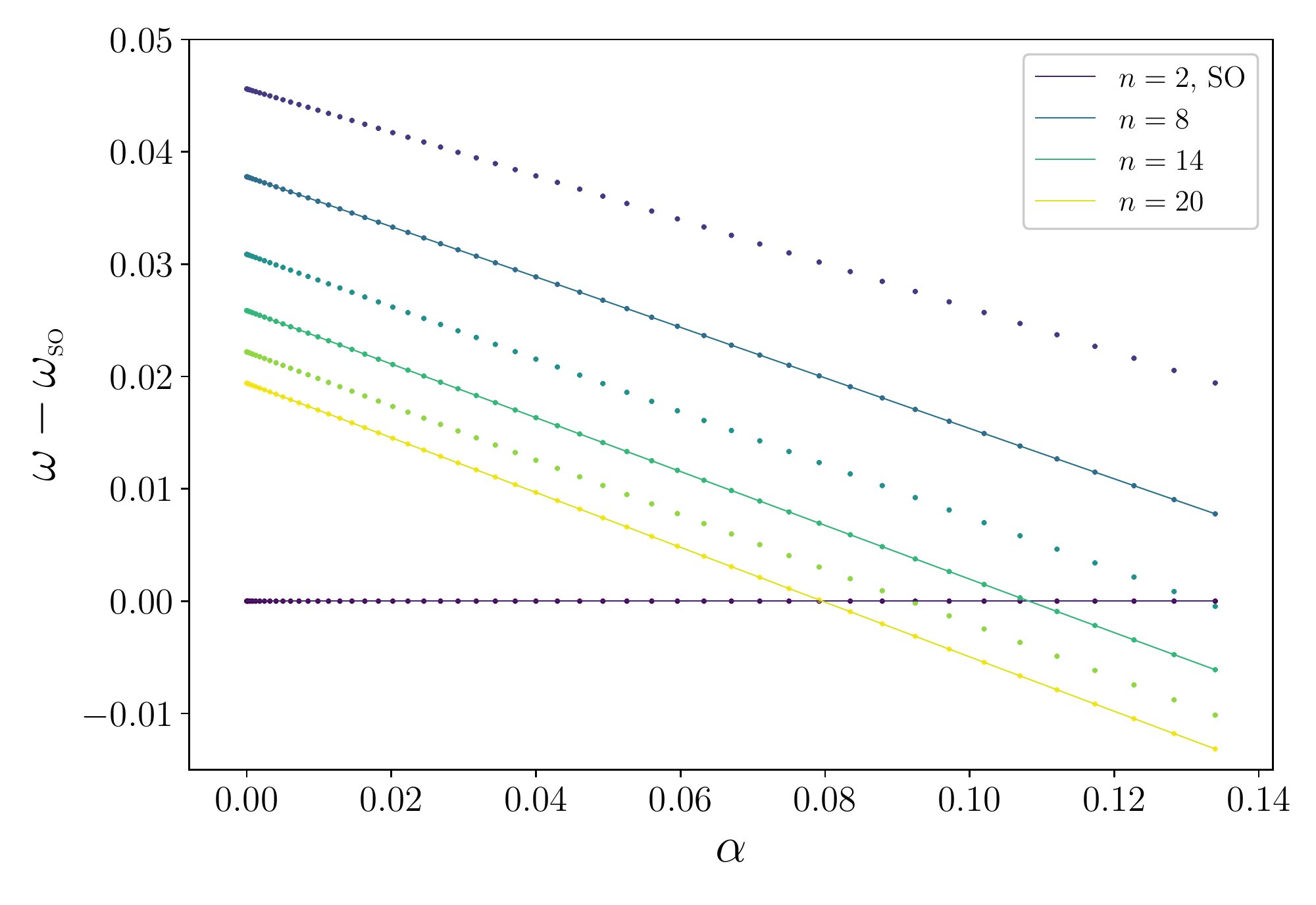}}\\
\caption{The spin-over mode in a spheroid can interact with nearby modes if the flattening is large enough. Panel (a) shows the distribution of theoretical (inviscid) eigenfrequencies at $\alpha=0.0202$), the red dashed line indicating the spin-over ($n=2$) mode frequency. Panel (b) shows the distance in the frequency axis between the spin-over mode and other nearby modes (antisymmetric ones indicated by continuous lines) as the flattening is varied. For details on the ensuing viscous interaction see  \citet{schmitt2006}.}
\label{fig:schm}
\end{figure}

By allowing the mantle to react to torques exerted by the fluid we extend the spectrum of purely inertial eigenmodes to include the Chandler Wobble (CW), which induces a flow in the bulk of the fluid core, however small in magnitude compared to the mantle's motion. Additionally, the Free Core Nutation (FCN) appears as an extension of the spin-over mode as soon as the mantle's moment of inertia ($1/q$) is considered finite. The FCN has a smaller mantle/fluid kinetic energy ratio than the CW, while other modes have even smaller ratios. As $q$ is varied the FCN may interact with the $n=8,14,20,\dots$ family of modes when they become close enough in the $(\omega,\sigma)$ plane, in a conceptually similar way as they interact with the spin-over mode as shown by  \citet{schmitt2006}, the difference being that in our case the relevant parameter is $q$ while in Schmitt's case it is the flattening. Figure~\ref{fig:schm} helps understanding why these modes are prone to either kind of interaction. In Fig.~\ref{fig:schm}(a) we show the inviscid eigenfrequency distribution for modes with azimuthal wave number $m=1$ for a spheroid with flattening $\alpha=0.0202$. We see that the equatorially antisymmetric $n=8,14,20,\dots$ modes are indeed the closest in frequency to the spin-over mode. Note that we can always find modes arbitrarily close in frequency to the spin-over mode, given that the inviscid spectrum is dense. However, those modes will have very complex structure, i.e. they will have high $n$ numbers, and would be very damped in the viscous case preventing them effectively from any interaction with the spin-over mode. In our case, with the FCN replacing the spin-over mode, increasing $q$ from zero causes the FCN frequency to go up at the same time that the damping increases in magnitude, resulting in complex interactions as shown in Fig.~\ref{fig:main}. In Schmitt's case, increasing the flattening causes the spin-over frequency to cross with the $n=8,14,20,\dots$ modes frequencies as well, although their damping is affected somewhat differently, \citep[see][their Figure 4]{schmitt2006}.

Figure~\ref{fig:schm}(a) also explains why the CW does not interact with any nearby modes. The only modes with small enough prograde frequencies (i.e. negative in our convention) like the CW, and possibly small enough damping, are equatorially symmetric, thus unable to interact.

Nothing stops both effects from acting simultaneously if the mantle is free to react to torques and the flattening is $\alpha\gtrsim 0.08$, in which case an astrophysically small Ekman number would not prevent the FCN to interact with the other modes as it does in our case. This is indeed an interesting perspective, however such a large flattening is beyond our reach given the Taylor expansion method that we use to implement the spheroidal boundary condition.

The width of the viscous boundary layer scales as $E^{1/2}$, so we expect that our Taylor expansion method is valid as long as the flattening is such that $\alpha/E^{1/2}\ll 1$. This might help a bit in understanding the somewhat surprising scaling of the FCN's $m_{2,\omega}$ and $m_{2,\sigma}$ coefficients. We can rearrange slightly the expression for the FCN's $m_\omega$ given by Eq.~(\ref{eq:omega_fcn}) as
\begin{equation}
m_\omega \approx E^{1/2}\left[ 0.599\,E^{0.043} + \left(\frac{\alpha}{E^{1/2}}\right) -0.427 \left(\frac{\alpha}{E^{1/2}}\right)^2 E^{0.001}\right].
\end{equation}
We see that the powers of $\alpha/E^{1/2}$ within the square brackets form a second order power series with coefficients that are either independent of $E$ or decrease slightly as $E$ decreases. Clearly, this series would not converge if the flattening is comparable or larger than the width of the viscous boundary layer.

The damping of the FCN presents a similar but slightly milder issue. The coefficient $m_{\sigma}$ shown in Eq.~\ref{eq:sigma_fcn} can be written
\begin{equation}
m_{\sigma} \approx E^{1/2}\left[ -2.705\,E^{0.002} + 11.58\,\left(\frac{\alpha}{E^{1/2}}\right)\,E^{0.553} +0.088 \left(\frac{\alpha}{E^{1/2}}\right)^2 E^{0.13} \right],
\end{equation} 
which shows the coefficient of $\alpha^2/E$ within the square brackets decreasing faster as $E$ decreases compared to the same coefficient in $m_\omega$'s case, i.e., $E^{0.13}$ vs. $E^{0.001}$. These `anomalous' scalings effectively prevent us from extrapolating the FCN's frequency and damping to larger flattenings. On the other hand, the Chandler Wobble appears free of such a complication and there is in principle no obstacle to extrapolate Eqs.~(\ref{eq:omega_cw}) and (\ref{eq:sigma_cw}) to higher values of the flattening. This is also true for the FCN's $b_\omega$ and $b_\sigma$ (that reduce to the spin-over mode frequency and damping respectively) where the scalings we found are not restricted to small $\alpha$, of course up to where Schmitt's mode interactions start becoming noticeable.

\begin{figure}
\centering
\subfloat[]{\includegraphics[width=0.5\textwidth]{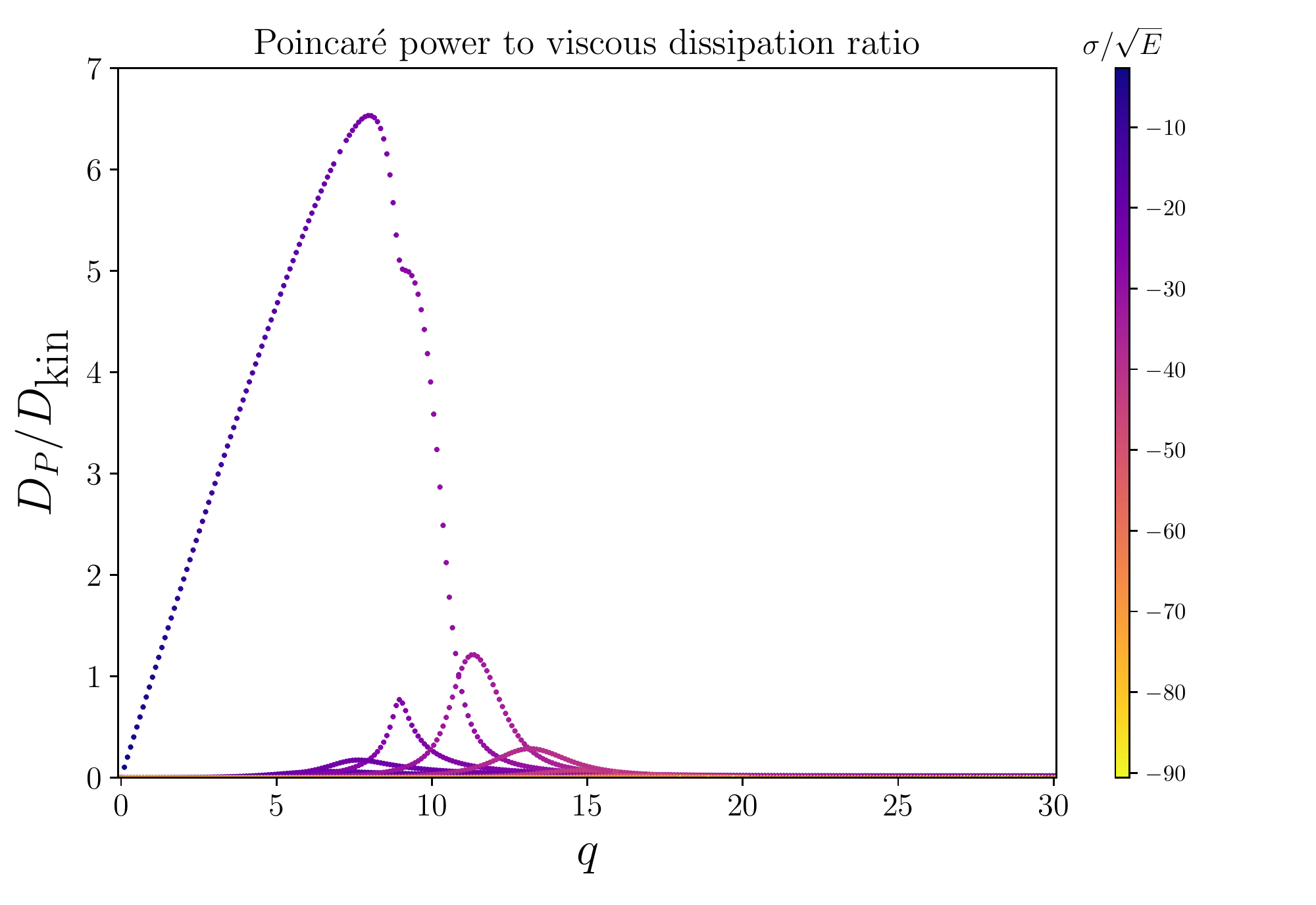}}
\subfloat[]{\includegraphics[width=0.5\textwidth]{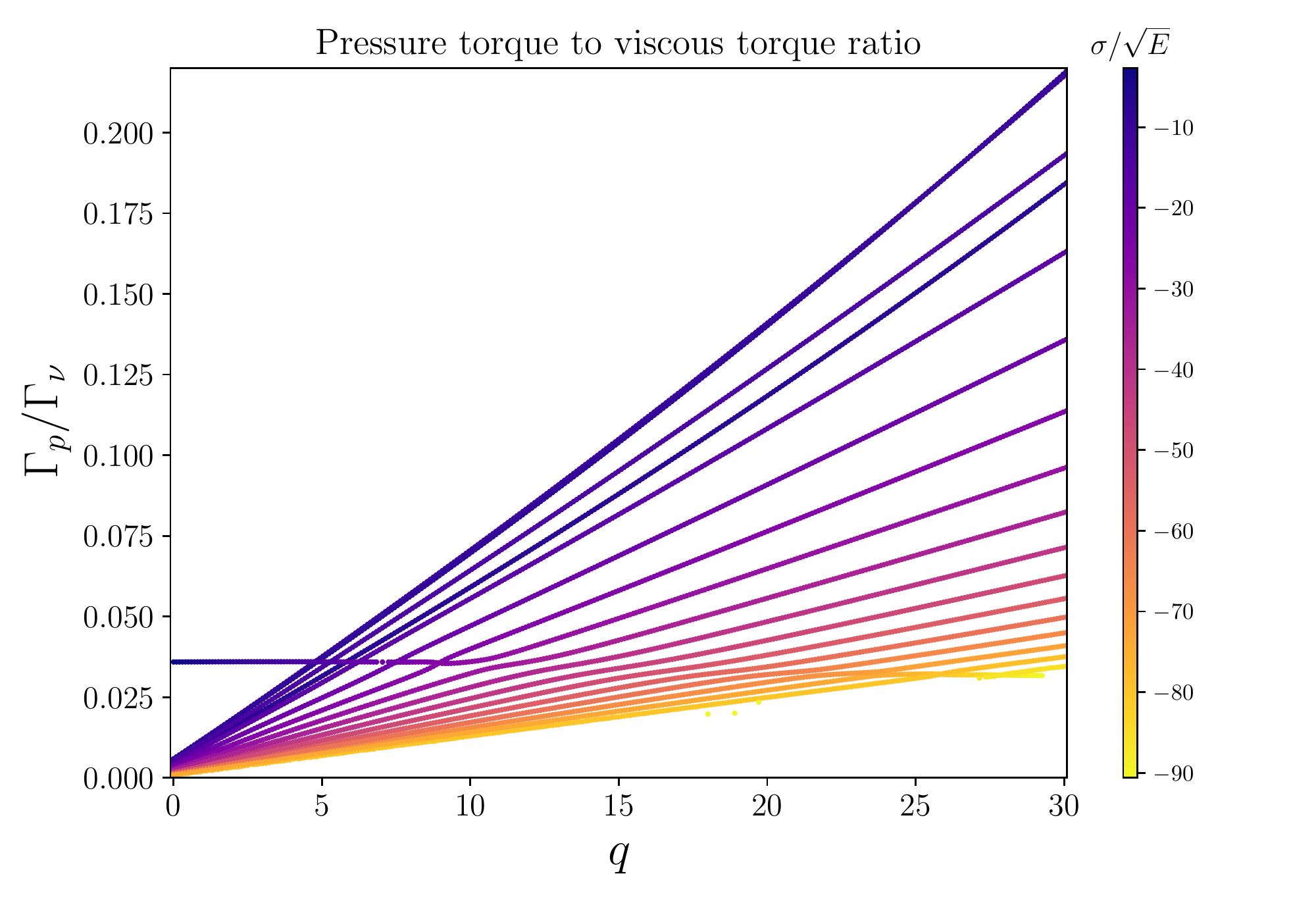}}
\caption{In panel (a) we can see that the power $D_P=|\partial_t\vb{M}\cdot\vb{L}|$ associated with the Poincar\'e force can be many times larger than the viscous dissipation $D_\text{kin}=|\mathcal{D}_\nu|$, thus dominating the overall damping rate. The pressure torque to viscous torque ratio in panel (b) shows that the viscous torque is dominant over the range of $q$ that we explored ($0\le q \le 30$). In both panels the parameters are identical to those of Fig.~\ref{fig:main}, color scale indicate the scaled damping $\sigma/\sqrt{E}$.}
\label{fig:ratios}
\end{figure}

The magnitude increase of the FCN's damping rate with $q$ is closely related to the fact that not all the damping is due to viscous dissipation processes. In fact, the Poincar\'e force is capable of performing work (as opposed to the Coriolis force) and such work must be taken into account in the overall power balance of the system. Starting from the linear momentum balance in Eq.~(\mbox{\ref{eq:ns}}), taking the dot product with $\vb{u}$ and integrating over the whole fluid volume it is straightforward to show that the damping rate $\sigma$ of any given mode fulfills
\begin{equation}
2\,\sigma\,K_f = \mathcal{D}_\nu - \left(\partial_t \vb{M}\right) \cdot \vb{L},
\end{equation}
where $K_f$ is the total kinetic energy of the fluid (in the mantle frame), $\vb{L}$ is the total angular momentum of the fluid (relative to the mantle frame) and $\mathcal{D}_\nu$ is the viscous dissipation computed as
\begin{equation}
\mathcal{D}_\nu=E\int \vb{u}\cdot \nabla^2\vb{u}\dd{V}
\end{equation}
over the whole fluid volume. For the FCN, as $q$ increases the mantle's motion also goes up (because its moment of inertia decreases), see Fig.~\ref{fig:main}(c), along with the ratio $|\partial_t\vb{M}\cdot\vb{L}|/|\mathcal{D}_\nu|$ as shown in Fig.~\ref{fig:ratios}(a). Near the maximum at $q\sim8$ the power associated with the Poincar\'e force is almost seven times as large as the power associated to viscosity. Beyond this value of $q$ mode interactions dominate and the FCN loses its character as a mainly rotational (toroidal) motion.

We should remark that the mode coupling we witness in this system is essentially viscous in nature for the sole reason that pressure torques exerted on the spheroidal mantle are much smaller than the viscous torques. Figure~{\ref{fig:ratios}}(b) shows that pressure torque to viscous torque ratio sits near 4{\%} when the mode interactions take place. That being said, it is the mixing among poloidal and toroidal components of different angular degrees induced by the torques that is at the heart of the mode interaction mechanism. However different in magnitude, both viscous and pressure torques induce similar mixing among components and so there is nothing in principle preventing pressure and gravitational torques, if they were to be dominant over viscous ones, from leading to similar resonant coupling among modes. This would be the case for a more realistic Earth model, where the Ekman number is very small and the pressure torques dominate. Note that additionally, a relatively large flattening such as Earth's, would bring the FCN's mode frequency closer to the region where it might interact resonantly with neighbouring modes regardless of the pressure torques, just as is the case for the spin-over mode as shown by \mbox{\citet{schmitt2006}}. Unfortunately, our Taylor expansion method prevents us from exploring such regime.

Lastly, other dissipation mechanisms such as Joule heating (in an electrically conducting core) might lead to similar resonant interactions. In such case magnetic torques would play a similar role as the viscous torque. This is particularly relevant for the Earth where we know that pressure, gravitational and magnetic torques dominate over viscous torques. It is then perfectly possible that resonant mode interaction, conceptually similar to the one presented in this study, might take place using more realistic (Earth or planetary) parameters.

\section{Summary and outlook}
\label{sec:end}
We have implemented an efficient spectral method  \citep{olver2013} to obtain numerically the eigenmodes of a rapidly rotating two-layer planet model in which the rigid mantle is free to react to torques exterted by the viscous, incompressible fluid core. The aim of this model is to understand at least qualitatively the relationship between inertial modes as studied in the fluid dynamics literature, where the rotational motion of the container (the `mantle') is prescribed, and the rotational modes of a planet (Free Core Nutation, Chandler Wobble) as studied in Geodesy where the motion of the fluid core is generally taken as a simple Poincar\'e flow. We can accomodate a flattened core-mantle boundary by means of a Taylor expansion technique as long as the flattening is smaller than the typical width of the Ekman boundary layer. In the special case of $q=0$ (i.e. a steadily rotating mantle) our model reproduces satisfactorily the viscous inertial eigenmodes in rotating spheroids as reported by previous studies  \citep{schmitt2006,zhang2004}.

When $q\ne0$ we find a rich system where the Free Core Nutation, which reduces to the spin-over mode when $q=0$, interacts with a family of nearby modes ($n=8,14,20,\dots$) and induces a substantial reordering. This family of modes that interact with the FCN include, incidentally, the three modes that  \citet{schmitt2006} describes as entering in resonance with the spin-over mode for large enough values of the flattening. We thus uncover a common underlying mechanism in which a varying parameter can bring modes' eigenvalues close enough in the complex plane, i.e. the $(\omega,\sigma)$ plane, leading to avoided crossings. If a crossing in frequency occurs, an avoided crossing in the damping takes place and vice versa.

The spin-over mode transitions into the FCN mode, increasing its frequency and becoming more damped as $q$ is increased. The other modes in the $n=8,14,20,\dots$ family are basically immune to changes in $q$ as long as the frequency and damping of the FCN is still away from the frequency and damping of any member of that family. The FCN, as $q$ keeps going up, gradually morphes into one of the members of the family, losing its own identity. The mode the FCN morphes into, with its identity taken away by the FCN, transforms itself into the next more damped mode of the family, and so on `bumping' successively each other into more negative values of $\sigma$. A clear explanation for this remarkable reordering still evades us.

The Chandler Wobble, just as the FCN, is affected by varying $q$, with the difference that the CW is not able to interact with other modes, the reason being that the very small damping of the CW is a difficult match for any low-order, equatorially antisymmetric modes with similar frequencies. The only modes that could match the CW's damping and frequency are low-order equatorially symmetric ones, but their interaction is precluded by their differing symmetry.

Away from the interaction region, when $q\lesssim0.1$, the FCN's frequency and damping seem to be approximatly linear with $q$. In the case of the CW, the frequency and damping appeared directly proportional to the flattening $\alpha$. We established the frequency and damping of both modes in this regime on the three control parameters $q,\alpha$ and $E$. We find that the theoretical inviscid expectation for the FCN frequency, to first order in $\alpha$, is recovered when $E\to 0$ but only as long as $\alpha E^{-1/2}\ll1$. We recover as well the inviscid CW frequency when $E\to 0$ this time without restrictions on $\alpha$.

The model we present in this study can be readily modified to include magnetic fields, in which case magnetic torques and ohmic dissipation are likely to be important if the Ekman number is reduced  \citep{buffett2010,lin2017}. Additionally, the inclusion of an inner core together with gravitational torques will bring two new modes, namely the Free Inner Core Nutation (FICN) and the Inner Core Wobble (ICW). The mere presence of an inner core will also impact significantly the modes and will probably increase the computational cost.

We acknowledge that our study is far removed from Earth's parameters and therefore it is not able to address quantitatively the shortcomings of current nutation models in relation to observations. In a speculative realm, our model might be of some relevance in the dynamics of rapidly rotating neutron neutron stars where rigid crusts are very thin (and with low density compared to the dense fluid core) and thus their $q$ parameter might be larger than unity, leading possibly to resonant mode interactions. In a less exotic note, however, our work provides a proof of concept indicating that inertial modes are relevant for Earth's nutations and that their study, including an inner core and magnetic fields, is well worth the effort. These additional features will render our model much more realistic in relation to the Earth or other planets and moons. They are relatively straighforward to implement and will certainly add surprises to the rich dynamics we already discovered.

\begin{acknowledgments}
We would like to thank Rapha\"el Laguerre, Ping Zhu and J\'er\'emy Vidal for many useful conversations and suggestions. We thank as well Ankit Barik for proofreading the article and giving constructive feedback. Also we express our gratitude to the PETSc and SLEPc team for their prompt help using their solver. The funding for this research is provided by the European Research Council through the Advanced ERC Grant `Rotation and Nutation of a Wobbly Earth' (RotaNut) No. 670874.
\end{acknowledgments}

\bibliographystyle{gji}
\bibliography{biblio}

\appendix

\section{The spheroidal boundary condition}
\label{sec:apa}
A convenient way to represent the no-slip boundary condition on the CMB involves the components of the velocity field $\vb{u}$ along the canonical spherical basis vectors  $\bm{\hat \epsilon}_0,\,\bm{\hat \epsilon}_+,\,\bm{\hat \epsilon}_-$, which we define as
\begin{equation}
\bm{\hat \epsilon}_-=\frac{1}{\sqrt{2}}(\bm{\hat \theta}-i\bm{\hat \phi}),\,\bm{\hat \epsilon}_0 =\vb{\hat r},\,\bm{\hat \epsilon}_+=-\frac{1}{\sqrt{2}}(\bm{\hat \theta}+i\bm{\hat \phi}).
\end{equation} 
Then the velocity field, anywhere in the fluid core, can be written as
\begin{equation}
\vb{u}=\sum_{\ell}\sum_{m=-\ell}^{\ell}\left[ u_{\ell m}^- d_{m,-}^\ell(\theta)\, \bm{\hat \epsilon}_- + u_{\ell m}^0 d_{m,0}^\ell(\theta)\, \bm{\hat \epsilon}_0 + u_{\ell m}^+ d_{m,+}^\ell(\theta)\, \bm{\hat \epsilon}_+ \right]\text{e}^{i m \phi},
\end{equation}
where $d_{m \nu}^{\ell}(\theta)\text{e}^{i m \phi}$ with $\nu=0,\pm$ are the generalized spherical harmonics  \citep{phinney1973}. The components $u_{\ell m}^{\nu}$ are related to the poloidal and toroidal scalars introduced in Eq.~(\ref{eq:poltor}) through
 
\begin{align}
\label{eq:u_poltor}
u_{\ell m}^-  &=\sqrt{\frac{\ell(\ell+1)}{2}}\left[ P'_{\ell m}(r) + \frac{P_{\ell m}(r)}{r} - i\,T_{\ell m}(r) \right],\\
u_{\ell m}^0  &=\ell(\ell+1) \frac{P_{\ell m}(r)}{r},\\
u_{\ell m}^+  &=\sqrt{\frac{\ell(\ell+1)}{2}}\left[ P'_{\ell m}(r) + \frac{P_{\ell m}(r)}{r} + i\,T_{\ell m}(r) \right].
\end{align}
 
The spheroidal shape of the CMB, introduced in Eq.~(\ref{eq:cmb1}) can also be described using generalized spherical harmonics by noting that $Y_{\ell}^m(\theta,\phi)=d_{m,0}^{\ell}(\theta)\,\text{e}^{i m \phi}$. Using the following identity
\begin{equation}
\label{eq:c3j}
d_{m \nu}^\ell\,d_{0 0}^{2j}=\sum_{L=\ell-2j}^{\ell+2j}(-1)^{m+\nu}(2L+1)
\begin{pmatrix}
\ell & 2j & L \\
m    & 0  & -m
\end{pmatrix}
\begin{pmatrix}
\ell & 2j & L \\
\nu  & 0  & -\nu
\end{pmatrix}
d_{m \nu}^L \equiv \sum_{L=\ell-2j}^{\ell+2j} C_{\ell j}^{L m \nu}\,d_{m \nu}^L,
\end{equation}
where the parentheses represent Wigner $3j$-symbols, and considering terms up to third order in the flattening $\alpha$ we can write
\begin{equation}
\frac{\left(r_\textsc{\tiny CMB}-1 \right)^k}{k!}=\sum_{j=0}^3 a_{k j}\, d_{0 0}^{2j}(\theta)\,\text{e}^{i m \phi},\,\,(k=0,1,2,3),
\end{equation}
where the coefficients $a_{kj}$ correspond to the matrix elements of
\begin{equation}
\setlength\arraycolsep{7pt}
\begin{bmatrix}
	1                                        & 0                                        & 0                                & 0               \\
	-\frac{\alpha}{3} - \frac{\alpha^2}{5} - \frac{13\alpha^3}{105}  &  -\frac{2\alpha}{3} - \frac{\alpha^2}{7} + \frac{\alpha^3}{21}   & \frac{12\alpha^2}{35} + \frac{96\alpha^3}{385}   & -\frac{40\alpha^3}{231} \\
	\frac{\alpha^2}{10} + \frac{3\alpha^3}{35}               &  \frac{2\alpha^2}{7} + \frac{\alpha^3}{7}                & \frac{4\alpha^2}{35} - \frac{48\alpha^3}{385}     & -\frac{8\alpha^3}{77}   \\
	-\frac{\alpha^3}{42}                             & -\frac{5\alpha^3}{63}                            & -\frac{4\alpha^3}{77}                    & -\frac{8\alpha^3}{693}	
\end{bmatrix}
\end{equation}
so that $a_{kj}$ is given by the matrix element at row $k+1$ and column $j+1$. Note that with this particular parametrization of the CMB shape, the semimajor axis of the spheroid has unit length. 
Putting all together, we write the canonical velocity components \emph{at the CMB} as a Taylor expansion:
\begin{align}
\label{eq:u_taylor}
u^\nu|_\textsc{\tiny CMB}  &= \sum_{k=0}^3 \frac{\left(r_\textsc{\tiny CMB}-1 \right)^k}{k!}\,\left. \left(\frac{\partial^k}{\partial r^k}u^\nu  \right)\right|_{r=1} \\
&= \sum_{\ell,m}\sum_{k,j} a_{k j}\sum_{\lambda=-2j}^{2j} C_{\ell j}^{\ell+\lambda,m,\nu}\,\left. \left(\frac{\partial^k}{\partial r^k}u_{\ell+\lambda,m}^\nu  \right)\right|_{r=1}\,d_{m \nu}^\ell(\theta)\,\text{e}^{i m \phi},
\end{align}
where the coefficient $C_{\ell j}^{\ell+\lambda,m,\nu}$ is defined in Eq.~(\ref{eq:c3j}). We can now substitute $u_{\ell m}^\nu$ and its derivatives at $r=1$ using Eq.~(\ref{eq:u_poltor}). We perform symbolically all the steps described so far, algebraically very cumbersome, with the help of \classname{TenGSHui}, a symbolic tensor calculus package for Mathematica written by  \citet{trinh2018}. The end result for the boundary condition comprises then three expressions, one for each $\nu$ component $u_{\ell m}^\nu|_\textsc{\tiny CMB}$ involving the poloidal and toroidal scalars $P_{\ell m},\, T_{\ell m}$ as Chebyshev polynomials evaluated at $r=1$. These symbolic expressions need to be computed just once. When assembling the matrices for any particular problem, the Chebyshev polynomials (and their derivatives) at $r=1$ and the coefficients $C_{\ell j}^{\ell+\lambda,m,\nu}$ need to be computed numerically at run time. We evaluate numerically the Wigner $3j$-symbols themselves using a fast and accurate code developed recently by  \citet{johansson2015}.

\section{The pressure torque}
\label{sec:apb}
The pressure is not involved in the dynamics of the fluid core but it enters in the rotational dynamics of the mantle via pressure torques. These torques need to be expresed as functions of the poloidal and toroidal coefficients $P_{\ell m}^k$ and $T_{\ell m}^k$. We begin by noting that the reduced pressure $p$ introduced in Eq.~(\ref{eq:redp}) is related to the physical pressure $P$, to first order in $\vb{M}$, through
\begin{equation}
\label{eq:redp2}
p_0 = \frac{P_0}{\rho_f}-\left(\vb{\hat z}\cross\vb{r}\right)\cdot\left(\vb{M}\cross\vb{r}\right) =  \frac{P_0}{\rho_f} - \frac{r^2}{\sqrt{6}}\left( M_-\,Y_2^1 - M_+\,Y_2^{-1} \right),
\end{equation}
where $p_0$ and $P_0$ are amplitudes satisfying $p=p_0\,\text{e}^{\lambda t}$, $P=P_0\,\text{e}^{\lambda t}$ with $P_0=\sum_{\ell,m}\Pi_{\ell m}\,Y_l^m$.
It turns out that the pressure torque $\gamma_-$, introduced in Eq.~(\ref{eq:lio}), only involves terms with $\ell=2,4,6$ and $m=1$. With the help of \classname{TenGSHui}, we take the \emph{consoidal} $\ell, m$ component (i.e. along the direction of $\grad Y_l^m$) of each term in Eq.~(\ref{eq:ns}), substitute the reduced pressure $p$ by the expression in Eq.~(\ref{eq:redp2}) and then solve for the $\ell, m$ component of the physical pressure $\Pi_{\ell m}(r)$. The resulting expressions for $\ell=2,4,6$ and $m=1$ are
\begin{equation}
\begin{split}
\label{eq:Pi}
\Pi_{2 1}(r)= &\frac{\left(252 E-42 \lambda  r^2+98 i r^2\right)}{42 r^2}P_{2 1}+\frac{\left(-252 E r-42 \lambda  r^3+14 i r^3\right)}{42 r^2} P_{2 1}' \\
&\quad+ 3 E P_{2 1}''+E r P_{2 1}^{(3)}-\frac{16}{21} \sqrt{2} r T_{3 1}-\frac{r }{\sqrt{3}}T_{1 1}+\frac{r^2}{\sqrt{6}}M_-,\\
\Pi_{4 1}(r)= &\frac{\left(15400 E-770 \lambda  r^2+1617 i r^2\right)}{770 r^2}P_{4 1}+\frac{\left(-15400 E r-770 \lambda  r^3+77 i r^3\right)}{770 r^2} P_{4 1}' \\
&\quad+3 E P_{4 1}''+E r P_{4 1}^{(3)}-\frac{3}{14} \sqrt{15} r T_{3 1}-\frac{24}{55} \sqrt{6} r T_{5 1},\\
\Pi_{6 1}(r)= &\frac{\left(48510 E-1155 \lambda  r^2+2365 i r^2\right)}{1155 r^2}P_{6 1}+\frac{\left(-48510 E r-1155 \lambda  r^3+55 i r^3\right)}{1155 r^2} P_{6 1}' \\
&\quad+3 E P_{6 1}''+E r P_{6 1}^{(3)}-\frac{5}{33} \sqrt{35} r T_{5 1}-\frac{64}{35\sqrt{3}} r T_{7 1},
\end{split}
\end{equation}
where $P_{\ell m}=P_{\ell m}(r)$ and $T_{\ell m}=T_{\ell m}(r)$ and likewise for their derivatives, which are indicated by primes.
At this point we have all the ingredients to compute $\gamma_-$ up to third order in $\alpha$. With the help of \classname{TenGSHui} this is
\begin{equation}
\begin{split}
\frac{\gamma_-}{i \pi \alpha}= &-\frac{4}{105} \sqrt{\frac{2}{3}}\left(11 \alpha ^2+27 \alpha -42\right) \Pi_{2 1}+\frac{32}{693} \sqrt{5} \alpha (2 \alpha -11) \,\Pi_{4 1}+\frac{160}{429} \sqrt{\frac{14}{3}} \alpha^2 \,\Pi_{6 1}\\
&+\frac{4}{105} \sqrt{\frac{2}{3}} \alpha (11 \alpha -18) \,\Pi_{2 1}'+\frac{32 \alpha  (32 \alpha -11)}{693 \sqrt{5}} \,\Pi_{4 1}'+\frac{32}{33} \sqrt{\frac{2}{21}} \alpha^2 \,\Pi_{6 1}'\\
&+\frac{4}{21} \sqrt{\frac{2}{3}} \alpha^2 \,\Pi_{2 1}''+\frac{32}{693} \sqrt{5} \alpha^2 \,\Pi_{4 1}''+\frac{32}{429} \sqrt{\frac{2}{21}} \alpha^2 \,\Pi_{6 1}'',
\end{split}
\end{equation}
where $\Pi_{\ell 1} = \Pi_{\ell 1}|_{r=1}$ and analogously for the derivatives. Substituting Eq.~(\ref{eq:Pi}) into the equation above results in an expression (too long to be included here) involving the functions $P_{\ell m}$ and $T_{\ell m}$ as Chebyshev polynomials evaluated at $r=1$ as required.

\section{Kinetic energy formulae}
\label{sec:apc}
The kinetic energy of the mantle, relative to an inertial reference frame (IRF) can be computed as
\begin{equation}
K_m^\textsc{\tiny IRF}=\frac{\rho_m}{2\rho_f}A\left( \left|\Omega_x\right|^2 +\left|\Omega_y\right|^2\right)+ \frac{\rho_m}{2\rho_f}C\left|\Omega_z\right|^2  =  \frac{\rho_m}{2\rho_f}A\left|M_-\right|^2+\frac{\rho_m}{2\rho_f}C,
\end{equation}
where $\rho_m A$ and $\rho_m C$ are the mantle's moments of inertia around an equatorial axis and around the short axis of symmetry, respectively. The fluid velocity relative to the IRF, $\vb{u}_\textsc{\tiny IRF}$, can be computed from the velocity relative to the mantle frame $\vb{u}$ through
\begin{equation}
\label{eq:irf1}
\vb{u}_\textsc{\tiny IRF}=\vb{u} + \vb{\Omega}\cross\vb{r}=\vb{u} + \left(\vb{M}+\vb{\hat z}\right)\cross\vb{r}.
\end{equation}
We compute the kinetic energy of the fluid relative to the mantle frame as
\begin{equation}
K_f=\frac{\rho_f}{2}\int \vb{u}\cdot\vb{u} \dd{V}
\end{equation}
where we take $\vb{u}=\vb{u_0}\text{e}^{\lambda t}+\vb{u_0}^{\dagger}\text{e}^{\lambda^\dagger t}$, with $(^\dagger)$ indicating the complex conjugate. Then the kinetic energy in terms of the poloidal and toroidal scalars, assuming a \emph{spherical} CMB here for simplicity, is given by
\begin{equation}
\begin{split}
\label{eq:kf}
K_f = \text{e}^{2\sigma t}\sum_{\ell,m>0}  8\pi \frac{\ell(\ell+1)}{2\ell+1}  &\left\{ [\ell(\ell+1)+1]\int_0^1 |P_{\ell m}|^2 \dd{r} + \int_0^1 r\,P_{\ell m}'P_{\ell m}^\dagger \dd{r} \right. \\
&\left. \,  + \int_0^1 r\,P_{\ell m}P_{\ell m}^{\prime\dagger} \dd{r}  + \int_0^1 r^2\,|P_{\ell m}'|^2 \dd{r}    + \int_0^1 r^2\,|T_{\ell m}|^2 \dd{r} \right\}
\end{split}
\end{equation}
where $\sigma$ is the real part of the eigenvalue $\lambda$.

In order to compute the fluid  kinetic energy in the IRF, we first note that the term $\vb{M}\cross\vb{r}$ in Eq.~(\ref{eq:irf1}) can be written as
\begin{equation}
\label{eq:mr}
\vb{M}\cross\vb{r} = \curl \left[-\frac{r\,M_-}{\sqrt{2}}\,Y_1^1(\theta,\phi)\,\vb{r}\right]
\end{equation}
therefore to compute $K_f^\textsc{\tiny IRF}$ we simply add $-rM_-/\sqrt{2}$ to the toroidal $\ell=1,m=1$ component $T_{1 1}(r)$ in Eq.~(\ref{eq:kf}) and then add $\rho_f C_f/2$ to the resulting kinetic energy, with $C_f$ being the moment of inertia of the fluid core around its short symmetry axis.

The mantle's linear velocity $\vb{v}_\textsc{\tiny SRF}$ in the Steadily Rotating Frame (SRF), i.e. the frame rotating with the mantle's mean angular velocity, is
\begin{equation}
\vb{v}_\textsc{\tiny SRF}=\left(\vb{M}+\vb{\hat z}\right)\cross\vb{r} - \vb{\hat z}_\textsc{\tiny IRF}\cross\vb{r}
\end{equation}
where $\vb{\hat z}_\textsc{\tiny IRF}$ is the vertical unit vector attached to the IRF. In order to compute $\vb{\hat z}_\textsc{\tiny IRF}$ we need to express $\vb{\hat z}_\textsc{\tiny IRF}$ using mantle's coordinates, which we do with the help of the Euler angles $\alpha,\beta,\gamma$. The cartesian components of $\vb{\hat z}_\textsc{\tiny IRF}$ in mantle's coordinates turn out to be $(-\beta,\alpha,1)$, with the angles given by
\begin{equation}
\alpha = \frac{\lambda M_x + M_y}{1+\lambda^2},\quad \beta=\frac{\lambda M_y - M_x}{1+\lambda^2}.
\end{equation}
With this result, and in complete analogy with Eq.~(\ref{eq:mr}), we write
\begin{equation}
\label{eq:mr2}
\left(\vb{M}- \vb{\hat z}_\textsc{\tiny IRF} \right)\cross\vb{r} = \curl \left[-\frac{r\,M_-}{\sqrt{2}}\frac{\lambda}{\left(\lambda-i\right)}\,Y_1^1(\theta,\phi)\,\vb{r}\right],
\end{equation}
which allows us to compute both $K_M^\textsc{\tiny SRF}$ and $K_f^\textsc{\tiny SRF}$ in the same way as explained above for the IRF.

\end{document}